# The Cost of Lunar Landing Pads with a Trade Study of Construction Methods

Philip T. Metzger*[1] and Greg W. Autry[2]

[1] Florida Space Institute, University of Central Florida, 12354 Research Parkway, Suite 200, Orlando, FL 32826; philip.metzger@ucf.edu.

[2] Thunderbird School of Global Management, Arizona State University, 401 North 1st Street, Phoenix, Arizona 85004

*Corresponding author

Abstract

This study estimates the cost of building lunar landing pads and examines whether any construction methods are economically superior to others. Some proposed methods require large amounts of mass transported from the Earth, others require high energy consumption on the lunar surface, and others have a long construction time. Each of these factors contributes direct and indirect costs to lunar activities. To identify the most favorable construction method and to evaluate the overall price range, these disparate factors have been quantified in terms of cost and combined in a trade study. The most important economic variables turn out to be the transportation cost to the lunar surface and the magnitude of the program delay cost imposed by a construction method. The program delay cost is the incremental value of a lunar outpost that will be lost because of the delay imposed by the construction time, i.e., a "lack of opportunity cost." This study finds that the cost of a landing pad depends sensitively on the optimization of the mass and speed of the construction equipment, so a minimum-cost set of equipment exists for each construction method within a specified economic scenario. Several scenarios have been analyzed across a range of transportation costs with both high and low program delay cost assumptions. It is found that microwave sintering is currently the most favorable method to build the inner, high temperature zone of a lunar landing pad, although other methods are within the range of uncertainty. The most favorable method to build the outer, low temperature zone of the landing pad is also sintering when transportation costs are high, but it switches to polymer infusion when transportation costs drop below about $110K/kg to the lunar surface. Several additional sensitivities are identified: the thickness of the pads is important (baking pavers gains advantage over microwave sintering when the pad is thinner); reliability is not a major factor (the least reliable system requires about 50% additional development cost to achieve target reliability, but development costs are shown to be only a minor part of the overall costs); and the lunar program's launch cadence sets a practical limit on the economic benefit of faster construction. It is estimated that the Artemis Basecamp could build a landing pad with a budgeted line-item cost of $229M assuming that transportation costs will be reduced modestly from their current rate ~$1M/kg to the lunar surface to $300K/kg. It drops to $130M when the transportation cost drops further to $100K/kg, or to $47M if transportation costs fall below $10K/kg. Ultimately, landing pads can be built around the Moon at very low cost, due to economies of scale.

## 1. Introduction

The exhaust of a rocket landing on or departing from the surface of the Moon dislodges surface dust, sand, gravel, and rocks, and it accelerates much of this ejecta to very high velocities [1-9]. In the absence of atmospheric drag such debris will travel vast distances with no loss of energy. Assets exposed to such an effective sandblasting, including spacecraft, science instruments, habitats, and other infrastructure, may sustain significant damage [10,11]. The severity is a function of its proximity to the landing site since the density of the expanding debris field decreases with distance. In the case of extremely powerful rocket engines, it is possible that surface material may be accelerated to lunar orbital velocities, resulting in a cloud of high-speed particulate in the vicinity of the Moon for an extended period. Ongoing studies are seeking to understand whether the interactions of this dust with the Moon's, the Earth's, and the Sun's gravitational fields and electromagnetic forces in the solar wind might retain it in cislunar space long-term so that it accumulates through the course of many lunar landings [12]. These are highly undesirable outcomes that would surely impede or threaten lunar exploration and development.

Fully mitigating this problem will likely require the construction of landing pads, which provide an ejecta-free surface for landing and launching rockets. Many construction techniques have been proposed [13-35]. Some of them require large amounts of mass brought from Earth. Others require large amounts of energy to sinter or melt the soil (regolith). Some methods are very slow, and this will delay the value of subsequent surface operations. A trade study of construction techniques and an evaluation of the cost must include all these factors.

We present a method to trade landing pad construction methods using cost metrics for all these factors. The cost of materials brought from Earth includes the transportation cost per kilogram to the lunar surface. The energy expenditure is measured in terms of the full, lifecycle cost of energy systems that must be built, delivered, and operated on the lunar surface. The time required by a construction method imposes a "program delay cost." Each method requires the technology be developed, tested, and operated on the Moon, each with its own costs. In principle, when these considerations are quantified appropriately it will produce a fair comparison of the potential construction methods. Even if quantifying these costs is challenging, the process of doing so brings transparency to the assumptions and insight into the technology selection process.

## 2. Landing Pad Requirements

Prior work has identified a two-zone strategy for lunar landing pads, since the requirements for the pad material are different in each zone [36]. An inner zone close to the touchdown point must withstand the high temperature and pressure of the plume stagnation region directly under the rocket engines of the lunar lander through the final moments of landing. It must also withstand the high temperature and pressure of the shock event at engine ignition for launch as the lander departs from the Moon. This inner region is limited in radius to a few meters around the engine nozzle because the exhaust gases expand radially away from that point, dropping in both temperature and pressure [37]. However, the radially expanding gases are accelerating to high velocity and this can erode and eject soil from the surrounding area, so the outer zone must stop erosion over an even larger radius. The model developed here includes the inner and outer radii as user-selectable variables, and for illustration purposes this paper will use $r_{inner} = 12$ m for the radius of the inner pad and $r_{outer} = 27$ m for the radius of the outer pad. These numbers were adapted from van Susante and Metzger [36], which used $r_{inner} = 5$ m and $r_{outer} = 20$ m for the plume of a 40 t lander with engines low under the vehicle and clustered near centerline, but here 7 m has been added to each radius to account for a multi engine lander that has engines 2 m diagonally off centerline and an additional 5 m uncertainty (or margin) in the landing accuracy. The inner radius can be estimated by using the equations of Roberts [38], for example, for the plume gas temperature versus radius from the centerline at each timestep during the descent of the lander. The outer radius was determined by van Susante and Metzger [36] by using Roberts' equations to determine the distance from centerline beyond which the shear stress of the plume gas is below the threshold where any lunar soil can erode. The nominal landing pad is illustrated in Fig. 1

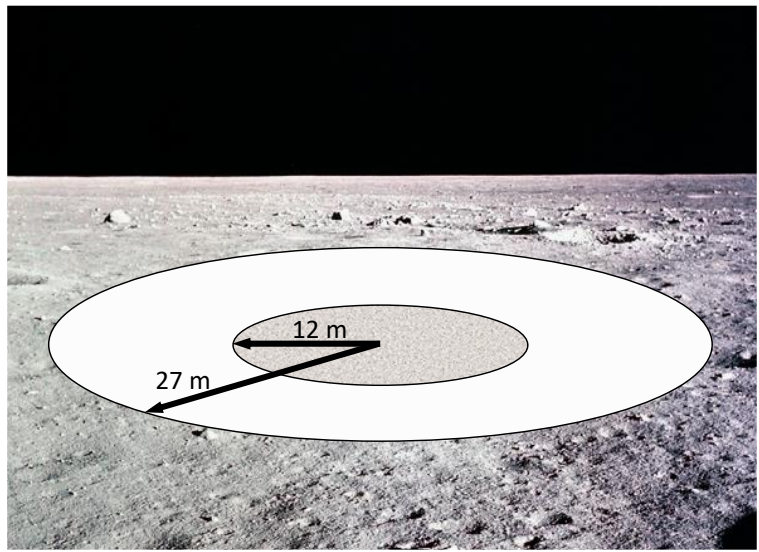

**Figure 1.** Landing pad configuration with inner and outer zone dimensions used in this study. (Lunar surface image credit: NASA)

## 3. Candidate Construction Methods

Many methods have been proposed to build landing pads, and this is an active area of research. Additional pad construction methods are proposed and studied every year. This paper will not attempt a survey of all possible construction concepts but will assess those that are presently better developed. We assume a landing pad will require site preparation (grading and compacting of the surface) as the first step in the construction process, followed by a soil stabilization method in the inner and outer zones.

*3.1 Inner Zone*

For the inner zone, this study assessed the application of microwave sintering and the use of pavers baked in an oven. Other possible inner-zone methods that are not considered here include the following. Solar sintering is promising [39-41] but work is needed to prove it will be effective for this application. Visible wavelengths have shallow depth of penetration in lunar soil so a pad must be built-up additively, and it remains to be shown that the resulting material will not delaminate and crumble when subjected to the thermal expansion and gas penetration of a plume's hot, high-pressure stagnation region. NASA used infrared sintering with a resistive heating coil to additively build-up a coupon in situ in volcanic tephra, but when subjected to a rocket thruster it delaminated and crumbled [18] so more work is needed to improve the process. Induction heating has been hypothesized for lunar industrial processes [42], and sintering by induction heating has been demonstrated for lunar soil simulant [43], but a literature search did not find a case of it being tested for applicability to landing pad construction. High temperature polymer has been infused in lunar soil simulant and has been demonstrated as an adequate heat shield for Mars entry [44]. This indicates polymer infused regolith might be a good material to use in the inner zone of a landing pad. It is not considered for this application here because it would impose a maintenance requirement after each landing to patch the ablated material, and it is not clear yet whether the maintenance needs will be excessive in cost or time or whether a single landing could burn all the way through the pad into the underlying soil resulting in catastrophic failure. More work is needed to answer these questions. A NASA Big Idea Challenge entry in 2021 proposed to use a fabric sheet over a polymer-infused base for the inner landing pad [45], but this was published too late to be included in this study and it will be assessed in future work. For now, polymer infusion is considered only for the outer zone of the pad where temperatures are low (see below). Fabric mats [46] or other flexible sheets [47] over unmodified soil have been proposed for landing pads. However, it is unclear whether fabric would block gas penetration adequately enough to prevent erosion of the underlying material that could compromise mechanical integrity of the pad. It is also unclear whether the soil anchoring method can adequately withstand the dynamics of plume gas diffusing through the fabric to build up an underlying gas pressure while the plume blowing at high velocity across the top of the fabric causes the Bernoulli effect. These methods and others might be competitors for economic pad construction but were omitted from the study due to technological immaturity and lack of knowledge how to successfully apply them.

Taylor and Meek [47,48] hypothesized that microwave sintering may be especially efficient for lunar materials since the fine lunar dust contains nanophase iron (np-$Fe^0$) particles in the glass patina that coats the grains. This patina with its np-$Fe^0$ is the result of space weathering processes on the airless lunar surface and was believed to make lunar soil more susceptible to microwave radiation. Recent research by NASA (Doug Rickman/NASA, personal communication) indicates that lunar soil with np-$Fe^0$ absorbs microwaves only marginally better than appropriate lunar simulants that do not contain np-$Fe^0$. Regardless, microwave sintering is attractive for construction because it is simple, without a lot of complicated robotics, and because the long wavelength of microwaves produces good depth of penetration in lunar soil, so the sintered material is sufficiently thick and mechanically competent after a single-pass construction process. This eliminates problems of delamination between layers from multiple passes. Work by NASA has shown that microwaved slabs are mechanically strong [14]. One challenge is the energy demand, since lunar soil has lower microwave susceptibility at lower temperatures as it is just beginning the heating process [50] so much of the microwave energy will pass through the sintering zone without being absorbed and thus be wasted. This will be quantified in this trade study.

Fabrication of pavers in ovens has been proposed as an energy-efficient alternative to in situ sintering of slabs. The oven keeps the energy contained while it diffuses into the center of the paver material, possibly reducing energy loss into the environment. This method was tested by Kelso et al. [51], including the manufacture of interlocking pavers from lunar soil simulant and their installation into a large-scale landing pad using a rover with a robotic arm. The economic challenges of paver baking include the extra mass of the ovens and the complex automation systems required to fill paver molds with soil and to remove and distribute the pavers after baking. The technological challenges include the extra robotic complexity making the hardware more expensive to develop and more expensive to maintain. Another challenge is keeping the plume gas in the high-pressure stagnation region from flowing through the cracks between pavers and building up a large area of high-pressure gas under the pad, which could result in catastrophic failure. Solving this might require grouting [52] to prevent gas intrusion, or grooves under the pad to allow the gas to flow back out rapidly and avoid pressure buildup. Care would need to be taken so that the grout will not mechanically fail under the thermal and mechanical loading of the plume. Nevertheless, we have long experience using bricks in terrestrial launch pads and the work to-date indicates this method can be successfully adapted to the Moon.

*3.2 Outer Zone*

For the outer zone, this study compares the use of polymer infused into the soil, spreading gravel and rocks obtained on the Moon, microwave sintering, and the use of pavers made in an oven. The polymer method is more easily applied in the outer zone than the inner zone since the temperatures will be low and thus the polymer will not break down during a landing event. The polymer-infused soil needs only resist the shear stress of the gas to prevent particulate erosion. This method of lunar construction has been demonstrated by [53]. It is very low-energy and very fast but introduces the economic challenge of the transportation cost of the many tons of polymer additive that must be brought from the Earth.

The method of using rocks or gravel to build up a "breakwater" structure for the outer landing pad has been innovated by van Susante [28]. Rocks can be raked from the regolith then sorted into different sizes using a trommel. In a reverse of the Macadam standard [54], the smallest sizes are laid on the ground first, with successively larger sizes on top. The rocks in the uppermost layer must be large enough that the direct action of the plume cannot lift them, while each lower layer is held in place by the one directly above it [55]. The pore spaces between rocks in each layer must be smaller than the rocks in the next lower layer to prevent them being pulled out by the gas. The gas that penetrates must be sufficiently slowed by the successively smaller pore diameters that it is unable to lift the dust and sand that lies beneath the lowest gravel layer. Unlike the polymer method, the gravel pad uses only in situ lunar materials so this may reduce transportation cost. However, the robotic mechanisms are more complicated than sintering or polymer infusion, and the construction times longer, resulting in a significant increase of development cost and program delay cost.

Microwave sintering and oven-baked pavers are the same processes for the outer pad as for the inner pad except that their thicknesses may be reduced because they do not need to withstand the downward pressure and the thermal stresses of the stagnation region of the plume. Making them thinner will save construction energy and time. For rover and foot traffic across the outer pad to and from the lander, a road may be constructed that is thicker than the rest of the outer pad to prevent fracturing it, but that detail was not included in this study.

## 4. Basic Assumptions and Physics-Based Modeling

The main focus of this paper is the economic analysis of landing pads, so the physics-based modeling of the construction techniques and the basic parameters of the trade study that underlie the economic analysis have been collected into Appendix A. These include parameters such as the energy used by rovers when grading, compacting, and constructing a lunar pad, the driving speed of a rover, the speed and energy consumption of rakes in lunar soil, the mass of microwave equipment as a function of microwave power, the energy flux needed to sinter lunar soil to a desired depth, the mass of polymer needed to infuse the regolith for a landing pad, etc. In each case a basis for estimation of the parameter has been documented from existing technologies, and where appropriate we used experimental data and physics equations for modeling the critical features of the construction methods. The model is necessarily based on estimations rather than measurement of mature lunar construction systems because each of these technologies is still under development. However, this approach does put reasonable limits on system performance, which enables us to satisfy the broad goals of the study. Documenting this model and its outcome also enables technologist to identify parameters to improve to make their construction methods more competitive and it will enable others to replicate and improve the model.

## 5. Non-Optimized (Initial) Results

The trade study model described in Appendix A was initially run without optimizing the size scale of any of the construction methods, using the hardware sets exactly as described in Table A-1. The model calculated the construction time, energy, and mass that must be brought from Earth for each construction method. These are listed in Table 1.

**Table 1. Initial Results for *Non-Optimized* Construction Methods**

| Construction Method | Time to Complete (days) | Energy Expended (MWh) | Mass from Earth (ton) | Number of Rovers | Max Power (kW) |
|---|---|---|---|---|---|
| Grading Inner Pad | 0.05 | 0.005 | 0.6 | 1 | 4 |
| Grading Outer Pad | 0.21 | 0.020 | 0.6 | 1 | 4 |
| Compacting Inner Pad | 0.10 | 0.011 | 0.5 | 1 | 4.3 |
| Compacting Outer Pad | 0.43 | 0.044 | 0.5 | 1 | 4.3 |
| Sintering Inner Pad | 4.1 | 19.7 | 4.7 | 4 | 200 |
| Sintering Outer Pad | 14.1 | 67.8 | 4.7 | 4 | 200 |
| Pavers for Inner Pad | 39.5 | 41.3 | 1.8 | 2 | 44 |
| Pavers for Outer Pad | 21.8 | 55.9 | 1.8 | 2 | 44 |
| Gravel/Rock Outer Pad | 35.4 | 0.81 | 1.6 | 2 | 1.0 |
| Polymer Outer Pad | 0.56 | 0.011 | 7.6 | 1 | 2.6 |

The mass brought from Earth is very high for sintering because this assumes 200 kW power will be expended to sinter as quickly as possible, and this necessitates 3.5 t of microwave hardware (magnetrons or other generators) plus 4 rovers to carry that mass of hardware. A cost-optimized sintering system will use fewer rovers and less mass of sintering hardware, taking longer to complete the construction. This economic optimization is discussed below.

Surprisingly, making pavers in the oven required more energy than microwave sintering for the inner pad, while making pavers took less energy than microwave sintering for the outer pad, even though the pavers and the sintered pad are specified to be the same thickness in each zone. That is because the oven loses more energy to the environment the longer it holds a constant temperature while heat slowly diffuses into the center of the pavers. Microwaves on the other hand penetrate the full sintering depth immediately due to the longer wavelength of microwaves compared to the shorter, thermal infrared (blackbody) radiation that does not penetrate lunar soil. There exists a particular paver thickness beyond which microwave sintering is more efficient than baking in an oven because of its speed. That crossover thickness appeared somewhere between 2.54 cm (outer pad thickness) and 7.62 cm (inner pad thickness) when using the parameters of this model.

For pavers in the inner zone, an estimated 84 kg of grout material is needed from Earth. It could be a sulfur-based or other waterless material that self-cures. If grout insertion proceeds 1 cm/sec along the length of the joints between the inner zone pavers, this adds another 55 hours to the construction process, 2/3 of the total construction time. Alternatives to grout include sintering soil in the crevices between pavers and rock welding between the pavers, but these methods require additional equipment, energy, and time. More work is needed to mature the grouting or alternative processes and understand the potential longevity of grout subjected to rocket exhaust.

In the first modeling attempt, the energy requirement for gravel/rock landing pads was higher than expected. It was dominated by the power needed to pull a rock rake through the soil fast enough to collect the rocks to build a pad in a reasonable amount of time. The energy estimate is highly dependent on the rock abundance of the local regolith because that determines the raking area and time. Because of the initial results, the method was modified to only sweep up rocks from the very top layer of the regolith rather than raking more deeply. This dramatically reduced the energy expenditure, which is shown in Table 1, although a larger surface area must be raked at the shallower depth. This change made the gravel/rock method far more competitive. Likewise, other innovations could change the competitiveness of any of the methods.

When combining the inner and outer pad construction techniques to build an entire pad, eight primary cases were considered as listed in Table 2. Each case also includes grading and compacting. When combining the construction times for each inner-outer pair, the construction processes were made serial rather than parallel for (1) grading, (2) compacting, (3) preparation of the inner pad, and (4) preparation of the outer pad. This is because the common rovers are needed for some of the processes and the mission control personnel performing landing pad construction are assumed to focus on one aspect at a time. The uplink/downlink data rate might also be constraining. For some construction methods the available power at the outpost might also be constraining. However, processes within the paver method (excavation of feedstock, hauling feedstock, baking feedstock into pavers, hauling pavers, and installing pavers) were parallelized since multiple rovers will be used in the optimized hardware sets and many batches will be performed for each process to complete the pad.

**Table 2. Eight Combined Cases of Pad Construction Methods**

| Signifier | Inner Pad | Outer Pad |
|---|---|---|
| SiSi | Sintered | Sintered |
| SiGr | Sintered | Gravel/Rock |
| SiPa | Sintered | Pavers |
| SiPo | Sintered | Polymer |
| PaSi | Pavers | Sintered |
| PaGr | Pavers | Gravel/Rock |
| PaPa | Pavers | Pavers |
| PaPo | Pavers | Polymer |

The construction energy for each of the eight primary cases is the sum of each process including grading, compacting, and stabilization of the inner and outer zones. The construction mass is the sum of the consumables (polymer and grout) and all the equipment required for all processes. When pavers or sintering is used in both the inner and the outer zones the equipment mass is included only once. Also, the rovers are used commonly used for all processes by swapping out the specialized attachments for grading, compacting, excavating, sintering, etc. The total number of rovers is decided by the process requiring the most of them, and the mass of rovers is not double counted when summing the masses for the processes. Using the input parameters from Table 1 without any economic optimization of the scale of each process, the model produced the following results. The total construction times are shown in Fig. 2. The consumed energies are shown in Fig. 3. The total masses brought from Earth are shown in Fig. 4.

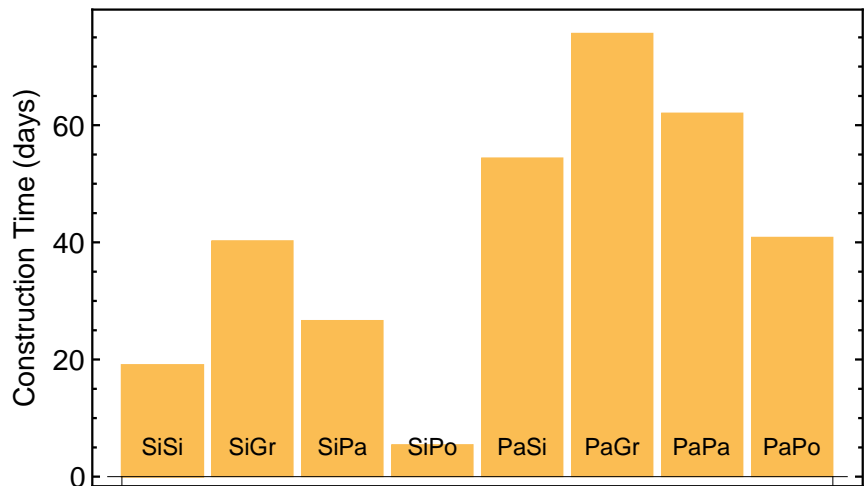

**Figure 2.** Construction times for the eight non-optimized cases.

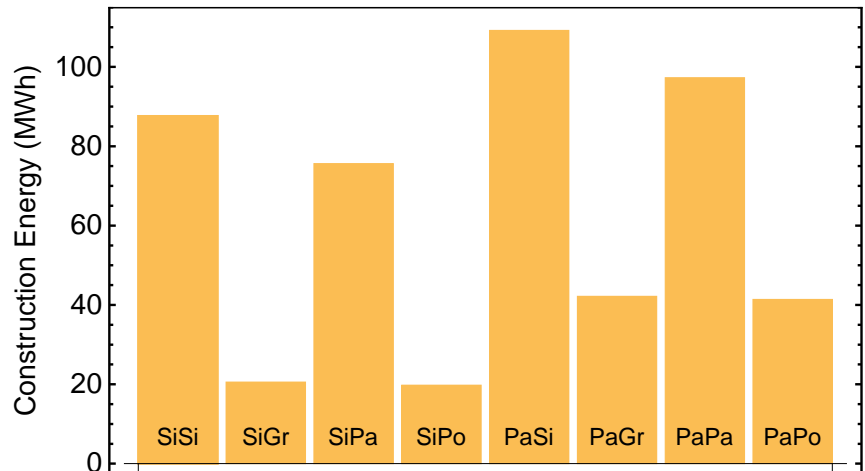

**Figure 3.** Construction energies for the eight non-optimized cases.

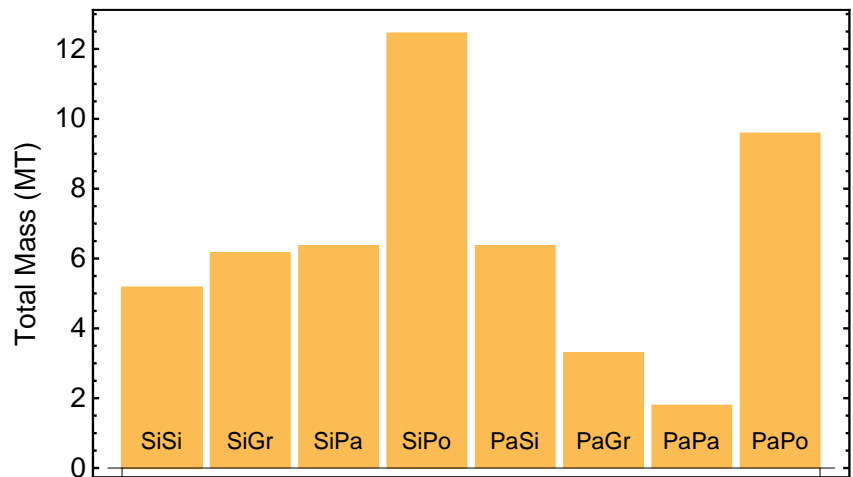

**Figure 4.** Total mass from Earth for the eight non-optimized cases.

We might naïvely think the most economical choice is PaPa since it requires the least mass from Earth. However, transportation cost is only one of the costs. Other costs include development cost, energy cost (because developing and transporting large energy systems to the Moon is a significant cost and commandeering a portion of the energy system's output takes away from other lunar activities that could use that energy, so this may be a real cost), and the program delay cost resulting from the time it takes to construct the pad reducing opportunities for the exploration and development objectives of the program. Because no construction method was the best in all three categories (Figs. 2, 3 and 4), this motivated the following economic model to merge all the parameters into a single cost metric. One of the purposes of this study is to evaluate the usefulness of this approach.

### 6. Non-Optimized Economic Comparison

The economic parameters in Table 3 were used to construct a single cost metric. They are user-selectable parameters in the model so other economic scenarios are easily tested.

**Table 3. Baseline Economic Assumptions**

| Parameter | Value | Units |
|---|---|---|
| *Program Data* | | |
| Budgeted Cost of the Lunar Program | 160 | $B |
| Estimated Value of Lunar Program (Low, Med, High) | 130, 520, 2600 | $B |
| Annual operational budget | 3 | $B |
| Program Duration | 20 | years |
| Discount Rate for Federal Money | 3.5 | % |
| Fraction of Program Reprogrammable (see text) | 75 | % |
| Hardware Development Cost Rate | 1.684 | $M/kg |
| Transportation Cost to the Lunar Surface | 300 | $K/kg |
| Yearly Operations Cost (for pad construction) | 124 | $M |
| *Energy Systems Data* | | |
| Solar Photovoltaic Mass-to-Power Ratio | 30 | kg/kW |
| Solar Photovoltaic Lifespan | 20 | years |
| Solar Duty Cycle | 80 | % |

*6.1 Cost of Program Delay*

The cost of program delay is estimated from the expected value of the program, which is estimated from the expected cost. The budgeted cost of the lunar program is assumed to be $160B capital investment plus $3B annual operating expense. This value was selected because they are roughly the development and operational costs of the International Space Station, and thus roughly indicates a known value that the U.S. Congress and international partners are willing to pay for an ambitious, international space project. Our goal is to estimate the order of magnitude of the value of a lunar program recognizing the large uncertainty. We will vary the parameters over a large range to make up for that weakness. Assuming $160B in equal allocations over a 20-year life of the program at the stated discount rate, the present value of those payments is calculated at $117.68B. However, we should assume that Congress, like other investors, expects a return on risky investments that exceeds the cost of the investment. Silicon Valley venture capitalists expect 10x returns on their winning investments in 5 to 10 years. Even conservative bankers expect returns that double their investments in a decade. A more realistic midrange estimate for the value of a lunar program would be 4x over 20 years or
$4 \times \$117.68B = \$470.7B$ present value of the program, and a high-end estimate would be at least 20x or $2.35T present value. One could argue that a 20-year return rate would be 10x each decade or 100x, so even 20x is conservative. Resource projections [56-58] suggest that the opportunity cost of "losing the Moon" to either non-development or to competing international coalitions (China and Russia are aggressively competing with Artemis) is very real and far exceeds this high-end value. We use 1x, 4x, and 20x in this study to conservatively estimate the total program value and the cost of delay, and we use 4x as the baseline case.

Construction of a landing pad must obviously precede all future activities requiring availability of the pad, such as large rocket landings close to an outpost and the delivery of resources dependent on a pad. Human Landing Systems are included in this category. Since Artemis is, by definition, a human lunar surface program, the program's intended value is primarily delivered by those surface operations. Any delay associated with landing pad construction postpones these programmatic activities and may generate an opportunity cost approaching the full value of program operations. Delaying the entire program at baseline value for one year at the discount rate would be $0.035 \times \$382.46B = \$16.48B$ in program delay cost. However, the entire program is not likely to be delayed by the landing pad construction time since other development activities on Earth and in space will be scheduled simultaneously. Landing pad construction may not be a critical path element at every moment. Nevertheless, program planners always choose shorter over longer construction times, other things being equal, offering prima facie evidence that program delay cost is a real factor in programmatic decisions.

We surveyed seven experienced program and project managers in the space industry to better understand the degree to which overall program delays can be mitigated when there is a "long-pole" task on the critical path. They stated that it depends strongly on the specific circumstances especially how far in advance the expected delay was identified. Without adequate anticipation, the overall program delay can even exceed 100% of the task delay due to the incompatibility of facility needs and employee skills for the parallel tasks, and due to the cost of retaining personnel with specialized skills who will be underutilized during the slowdown. However, a good lunar program management effort would endeavor to re-sequence the missions, and alternate approaches might be pursued in parallel to see if one may be completed before another, so the team might mitigate an unanticipated 1-year critical path delay down to just nine months, while an extremely good outcome might be six months. Tory Bruno, CEO of the United Launch Alliance (Tory Bruno, personal communication, 2022) wrote,

> A good rule of thumb is that a solid 2/3s of other tasks can proceed unaffected by the critical path item's delay. A really good program management team will have identified potential delay risks in advance and have protected the means to continue progress in the event that a risky item gets stuck and becomes critical path…A team that is experienced, and does risk and opportunity management aggressively, can often push the number of tasks that continue to 75%, under ideal circumstances.

In this trade study, we assume that a lunar program will determine the requirement for landing pads early enough to optimally mitigate 75% of the delay. The baseline program delay cost is estimated as 25% of the program's present value multiplied by the discount rate multiplied by the delay time.

The delay time is the construction time divided by the solar duty cycle. For example, the PaPo (non-optimized) construction time is 40.85 days (0.112 year). That assumes surface operations throughout the month, although for a solar-powered outpost any operation going beyond lunar sunset (every 29.5 days) must be delayed until the next sunrise. We assume the entire lunar surface schedule is affected on average by the solar duty cycle, so it is applied proportionately to all tasks. A human-tended outpost might use a nuclear fission reactor [59] instead of solar, but early construction tasks like building landing pads might be performed before the nuclear system is fully delivered and activated. One of the standard concepts is to set the nuclear reactor into an excavated pit some distance from the outpost to shield the crew from neutron radiation then deploy power cables across the distance. These tasks with the projected excavation rates may take 50 days to complete to complete [60] and will require solar power in the interim, and furthermore solar power may be retained as backup power even after the nuclear system is complete. For this study we assume a solar power system will be used for the initial landing pad construction.

The solar duty cycle may be 80% at well-lit locations near the lunar poles or 50% near the equator. The construction time of PaPo (in the non-optimized construction set) is 40.85 days, so dividing by 80% duty cycle and reducing the delay by 75% via good program management the program delay cost comes to $576M. The SiPo (non-optimized set) construction time is 5.45 days (0.015 year) so its program delay cost comes to only $77M, recovering $499M in program value compared to PaPo. Program managers may disagree on the values to use, but there must be some penalty for a slower landing pad construction method, and this economic approach provides a framework to rationally assess it. The parameter values are varied over a wide range, below, to test the sensitivity of the economic metric and the meaningfulness of the program delay cost.

We note also that incrementally decreasing the construction time might not incrementally recover program delay cost, because the launch cadence of lunar missions sets a practical limit on how fast surface activities can proceed. The launch cadence is determined by higher-level programmatic, budgetary, and policy decisions. The calculated program delay cost in this trade study is simply a metric, and like all trade studies it is not designed to be followed blindly but provides insight to inform those higher-level decisions. This will be discussed further toward the end of the paper.

*6.2 Hardware Development Cost*

The baseline development costs of hardware are based on the Global Security cost estimating tool [61], which we used to create a dollars-per-kilogram parameter. This parameter was applied to the mass of hardware in each construction method (but not to the consumable polymer mass and grout) and to the mass of solar power systems in calculating the value of the energy consumed.

*6.3 Transportation Cost*

Estimates of the transportation cost to the lunar surface vary widely depending on whether one is an optimist or pessimist about future prices. In the next five years they are widely expected to be on the order of $1M/kg. Within 10 years they may be somewhere in the wide range of $1M/kg down to $100K/kg and within 20 years perhaps $500K/kg down to $2K/kg. Our baseline economic scenario uses $300K/kg for a pad built in the 5-to-10-year timeframe. This is varied over a wide range to test sensitivity and meaningfulness of the results.

We have ignored the cost of the risk of launch failures during transportation of equipment to the Moon. Planning for and recovery from a launch failure require higher-level program decisions. We can assume the cost of this risk is offset by the increased value of the delivered assets when they are disposed of at the end of pad construction. I.e., a functioning construction rover on the lunar surface is more valuable than a new one on the Earth, because it has been "de-risked" and its value will be transferred to other programmatic functions or it will be sold to a commercial operator, in situ.

*6.4 Cost of Reliability*

Factors that affect reliability include (1) complexity, (2) state of the art, (3) performance time, and (4) operating environment. The construction methods studied here vary widely in these four factors. Reduced reliability increases cost by necessitating greater provisioning of spares and causing additional cost of delay when failures occur, which could be extreme. It is not feasible at this stage to project spares provisioning and repair strategies, but we can quantify the cost of reliability by assuming all construction methods will be matured to the same high level of reliability, quantifying the additional hardware development cost this entails. This approach is consistent with the arguments of Jones [62] that relying on spares and repairs for space missions is not adequate, so building greater reliability into every element to directly achieve a specified reliability is preferred.

First, we project what the baseline reliability will be for each subsystem (rover, excavator, sintering apparatus, etc.) if it were built using equally reliable components with equal design resilience but subjected to their own conditions in the four reliability factors, above. This is done using the Feasibility of Objective Technique (Military Handbook [MIL-HDBK]-338B) [63]. This technique adds the relative failure rates of subsystems to obtain the relative failure rate of each overall construction technology, normalizing these relative rates to an expected absolute reliability. We normalized them such that the most reliable of the four construction technologies will have reliability $R = 99\%$. The calculations are shown in Appendix B and the resulting baseline reliabilities are listed in Table 4. This exercise indicated that paver fabrication will have the lowest baseline reliability due to the intricacy of the robotics that fill paver molds with regolith then transfer the baked pavers from the molds onto the rover for installation, the risk of granular flow jamming while conveying regolith from the excavator into the paver molds, and the risk of lunar dust degrading the many mechanisms. These are all solvable problems, but this indicates the cost of development must be higher to achieve equal reliability with the other construction methods.

**Table 4. Reliability Cost Factors**

| Construction Method | Baseline Reliability (%) | Reliability Cost Factor | Resulting Reliability (%) |
|---|---|---|---|
| SiSi | 97.85 | 1.31 | 99.00 |
| SiGr | 96.42 | 1.37 | 99.00 |
| SiPa | 74.57 | 1.46 | 99.00 |
| SiPo | 99.00 | 1.00 | 99.00 |
| PaSi | 90.21 | 1.48 | 99.00 |
| PaGr | 88.89 | 1.48 | 99.00 |
| PaPa | 68.75 | 1.51 | 99.00 |
| PaPo | 91.27 | 1.35 | 99.00 |

Second, the Minimization of Effort Algorithm of MIL-HDBK-338B is applied to determine how much each subsystem must be improved to minimize cost while achieving equal 99.00% reliability in each overall construction method. The calculations are shown in Appendix B.

Third, the costs to achieve these subsystem improvements are estimated using the cost-reliability model of Mettas [64],

$$c_i = \exp\left[(1-f)\frac{R - R_{\min}}{R_{\max} - R}\right]$$

(2)

where $c_i$ is the cost factor for the $i$th subsystem that is multiplied onto its baseline cost, $R_i$ is the target reliability of the subsystem as determined by the Minimization of Effort Algorithm, above, $R_{\max} \cong 100\%$ is the maximum achievable reliability of the subsystem, $R_{\min,i}$ is the baseline reliability of the subsystem if it were built using baseline quality components and baseline design resilience as determined by the Feasibility of Objectives Technique, above, and $f$ is a parameter between 0 and 1 that estimates the feasibility to improve the reliability above the baseline. Mettas refers to Kecedegliou [65] for considerations to estimate $f$ and refers to engineering judgement. Stancliff, et al. [66] applied this model to lunar rovers using $f = 0.5$ and $f = 0.95$ showing similar results in each case. Stancliff, et al. [67] applied it again to lunar rovers using $f = 0.95$. Here we use $f = 0.5$ and we show that reliability is not a strong determinant of the trade study; a larger value of $f$ would make it even less of a determinant. The overall cost factor for a construction system is then $c = \phi_1 c_1 + \phi_2 c_2 + \cdots + \phi_N c_N$ where there are $N$ subsystems and $\phi_i$ is the fraction of the system mass in the $i$th subsystem since our cost estimating model is based on hardware mass. These overall cost factors are multiplied onto the hardware development cost in each case, and they are shown in Table 4. Although the baseline reliabilities varied widely and the cost factor grows exponentially, the cost factors are all < 2 because only the least reliable subsystems require extra development cost to achieve overall parity in each case. As shown below, the hardware development costs turn out to be a minor part of the costs except when transportation cost is extraordinarily cheap (see Fig. 8), so we conclude that reliability does not play a significant role in the trade.

*6.5 Energy Usage Cost*

Energy cost is the prorated part of the full cost of delivering energy on the Moon. This presumes the solar power plant will be transferred to other programmatic utilization after pad construction or sold to a commercial operator, in situ. The energy systems development cost ($/kW) is calculated from the solar photovoltaic mass-to-power ratio multiplied by the same hardware development cost-per-mass factor from the Global Security cost estimating tool as described above. The delivery cost ($/kW) is the solar photovoltaic mass-to-power ratio multiplied by the transportation cost to the lunar surface. The sum of these is taken to be the present value after it has been delivered to the Moon, so the annual cost over the twenty-year solar photovoltaic lifespan is found by the annual payment/present value equation to be $4.14M/kW/year. Dividing this by the hours in a year, the energy cost in the baseline case is found to be $473K/MWh. The energy cost is less in scenarios where the lunar transportation cost is less.

We note that energy cost may not always be a "real" cost on the lunar surface because, unlike the terrestrial power grid that has hundreds of millions of users resulting in smooth statistics and smoothly proratable business decisions, the energy needs of lunar construction are singular and, in some cases, comparable to the entire energy budget of the outpost [68]. Choosing a lower energy construction method may not always result in a real reduction of cost for the program. Nevertheless, energy will be a real factor in programmatic decisions, so an energy metric is needed. We do not believe there is any perfect framework for trade studies because they are too complex, but we have proposed the economic approach (and are evaluating it in this paper) because we think it provides advantages. For the calculated energy cost metric these advantages include (1) driving optimization away from extreme power usage, (2) putting energy into the same evaluation framework as the other factors of the trade study, (3) avoiding potential bias by being quantitative and objective, (4) providing an economic evaluation that is important for future commercial ventures, and (5) providing insight to inform the higher-level programmatic decisions, which will be discussed further toward the end of the paper. Also, the results will show that energy is not a significant cost driver under current economic conditions, but it may become important in the future as launch costs are reduced and lunar demand increases with surface activity.

*6.6 Non-Optimized Systems, Baseline Economic Scenario*

The baseline economic scenario represents a landing pad built for the Artemis lunar basecamp. The parameters are as shown in Table 3 including transportation cost of $300K/kg to the surface and a program delay cost based on a $160B-expense surface outpost that will operate for 20 years with expected 4x value-over-investment, and 75% of program delay cost mitigated through proactive program management. Each of these costs was calculated for each construction case (non-optimized) and graphed in Fig. 5. The non-optimized costs are dominated by delivery cost and program delay cost. The other costs are barely visible on the graph. SiPo has high delivery cost but low program delay cost because the masses of polymer and microwaving hardware are high, but the construction time is very short. This will be optimized by reducing the mass of microwaving hardware at the expense of higher program delay cost to find the minimum sum of all the costs.

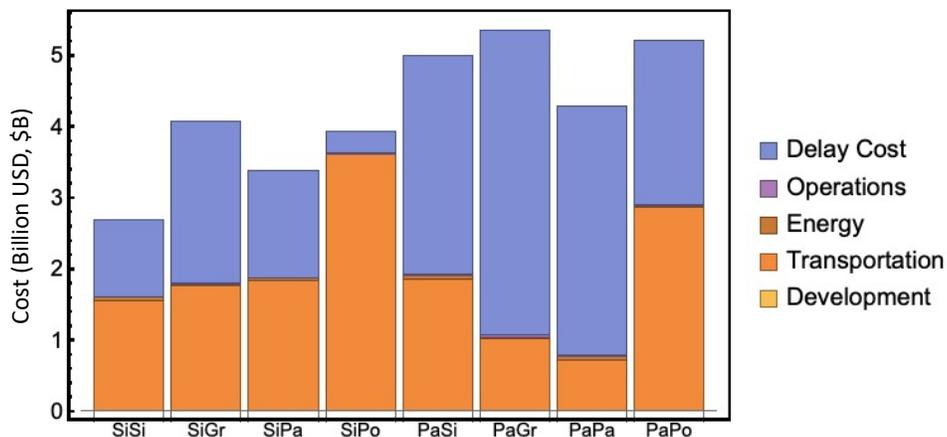

**Figure 5.** Costs of each landing pad construction scenario (non-optimized).

## 7. Optimizing the Construction Hardware

For each construction method there is an optimum mass of hardware for a given economic scenario. Each construction method can build a landing pad faster if it has more hardware working in parallel (more mass of hardware also resulting in higher power demand), which will increase the transportation cost but reduce the program delay cost. Three examples of the hardware scaling are shown in Fig. 6, demonstrating how a cost minimum exists in each case. The minima will shift left or right when any of the economic parameters change, so optimized scale of a construction set depends on the economic assumptions.

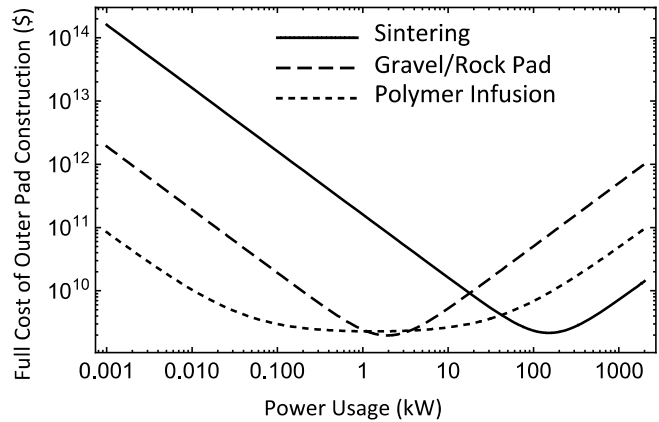

**Figure 6.** Three examples of scale optimization for outer pad construction methods. Each method's hardware construction set is scaled here according to the power it consumes. This is for the outer pad in the baseline scenario of Table 3.

The optimized costs for the baseline economic scenario are shown in Fig. 7. Some of them are significantly lower than the costs for the non-optimized systems in Fig. 5, demonstrating how scale optimization is important for a trade study. Only transportation and program delay cost are significant in this economic scenario, the other costs being invisible or barely visible on the plot. Methods that use polymer are not as competitive due to the high transportation of consumable polymer from Earth even though the fast application of polymer reduced its program delay cost. SiSi is the economically best method at $2.64B full cost. Since the program delay cost and energy costs would not be included in the appropriation line item for the landing pad, the "appropriated cost" subtracts those two cost elements and is $1.31B. This is still very expensive so political pragmatism to lower the appropriated cost will be discussed below.

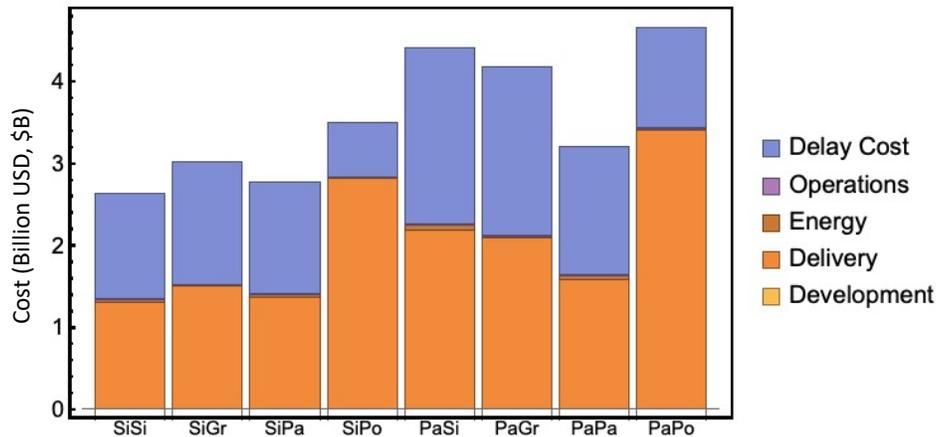

**Figure 7.** Optimized full-cost of pad construction for the pairs of construction methods in the baseline Artemis Basecamp case.

## 8. Varying the Economic Scenario

Six economic scenarios are examined to test whether other landing pad construction technologies than SiSi would be more favorable in different economic scenarios and to understand the sensitivity of the assumptions in this trade study. The results are shown in Fig. 8. These cases use either expensive, moderate, or cheap transportation and either high, moderate, or no program delay cost. Expensive transportation is $1M/kg to the lunar surface representing current conditions. Moderate transportation is $100K/kg to the lunar surface, which is a factor of 10 reduction from current costs. Many believe it is achievable after about a decade. Cheap transportation is $300/kg, which represents the more distant future when there is significant economic activity in space, and it tests the extremes of the model. High program delay cost is based on a $160B outpost with expected 20x return on investment and 75% of the delay cost mitigated. Moderate program delay cost is based on a $160B outpost with only 1x return on investment and 75% of the delay cost mitigated. The scenarios assuming there is no program delay cost are unrealistic since the high delivery cost militates against building landing pads unless they are needed for important surface activity, which implies the existence of program delay cost. However, it may be that simply doing activity on the Moon is the intended value (i.e., projecting national presence on the Moon) so completing the pad may be less urgent than simply being there to work on it. Also, the no-value scenarios demonstrate the model's behavior over the full range of conditions.

*8.1 Expensive Transportation/High Delay Cost Scenario*

The first case is shown in Fig. 8(A) with expensive transportation and high program delay cost. The expectation of high value drives the optimization toward rapid construction requiring high mass of equipment increasing the transportation cost. The result is roughly equal transportation cost and program delay cost, except for the construction sets that use polymer because the large consumable mass does not scale. Other costs are negligible. Methods that use the same technology for inner and outer pads (SiSi and PaPa) have a cost advantage due to commonality of hardware reducing transportation cost.

The most economical method is SiSi, which requires 5.3 t of hardware (including the rover attachments for grading and compacting). It takes 1.0 day for grading and compacting, 5.0 days to sinter the inner pad, and 17.3 days to sinter the outer pad. PaPa requires 6.3 t of hardware and takes 8.6 hours for grading and compacting (faster than SiSi since the cost-optimized hardware set for PaPa uses more rovers), 17.9 days to pave and grout the inner pad, and 9.9 days to pave the outer pad. The appropriated costs for SiSi, SiPa, and PaPa in this scenario are $5.28B, $5.56B, and $6.45B, respectively.

*8.2 Expensive Transportation /Low Delay Cost Scenario*

The second case in Fig. 8(B) uses expensive transportation and low program delay cost. All construction methods are now less expensive due to lower expectation of value driving it to smaller hardware sets with longer construction times. The use of polymer did not drop in cost as much as the other methods since the required quantity of polymer does not scale and still suffers from high transportation cost. For the least expensive SiSi method, the hardware mass is 1.2 t, grading and compacting take 4.3 days, sintering the inner pad takes 22.2 days, and sintering the outer pad takes 76.4 days. The appropriated costs for SiSi, SiPa, and PaPa in this scenario are $1.23B, $1.30B, and 1.57B, respectively.

*8.3 Expensive Transportation/No Delay Cost*

The third case in Fig. 8(C) uses expensive transportation and no program delay cost. In this scenario the use of polymer is not competitive due to the transportation cost while gaining no offsetting benefit from the faster speed of the method. On the other hand, this demonstrates the importance of including program delay cost in a trade study, because without it, polymer seems less viable than it is. Operations is the second highest cost category in this scenario because the high transportation cost drives the optimization toward a small set of hardware that increases the construction time and hence the operations cost. SiSi is still the least expensive approach with appropriated cost $409M and total equipment mass of 204 kg. It requires 25 days to grade and compact, 130 days to sinter the inner pad, and 447 days to sinter the outer pad for a total construction period of one year and eight months. This slow speed seems unrealistic, but that is because the absence of a program delay cost may also be unrealistic.

*8.4 Moderate Transportation/High Delay Cost*

The fourth case in Fig. 8(D) uses moderate transportation and high program delay cost. The reduced transportation cost enables polymer to be more competitive, and the high program delay cost favors the speed of the polymer method, so SiPo has the lowest total cost at $2.46B (appropriated cost $1.58B). Its appropriated cost is close to the appropriated cost of SiSi ($1.68B). SiPo builds the pad in a total of 3.8 days (14 hours grading and compacting, 3.0 days sintering the inner pad, and only 8 hours applying polymer in the outer pad) whereas SiSi takes a total of 7.4 days, so SiPo wins mainly by its speed.

*8.5 Moderate Transportation/Low Delay Cost*

The fifth case is shown in Fig. 8(E) with the moderate transportation cost and the low program delay cost. Unlike the fourth case, SiPo is not the most economical because the lower program delay cost no longer benefits as much from the faster application of polymer. However, all the methods are less expensive in this case than in the prior case. The most economical is SiSi with total cost $799M ($392M appropriated cost), building a landing pad in 32.8 days.

*8.6 Moderate Transportation/No Delay Cost*

The sixth case in Fig. 8(F) uses moderate transportation and no program delay cost. The use of polymer is not competitive. SiSi has the lowest total cost at $168M ($130M appropriated cost) building a pad in 192 days. SiPo saves 108 days of construction time but has an appropriated cost that is $656M higher than SiSi.

*8.7 Cheap Transportation/High Delay Cost*

The seventh case in Fig. 8(G) uses cheap transportation and high program delay cost. In all the cases with cheap transportation, SiPo is the most economical. In this case the total cost is $252M (appropriated cost $123M). The grading and compacting attachments are 6.2 tons of hardware operating on multiple rovers in parallel to complete their task in 2 hours. This is comparable to a terrestrial site preparation project. The system uses 57.6 t of sintering hardware including rovers to finish the inner pad in 10 hours. It uses 1.8 t of polymer application hardware (not including rovers) to apply the 7.2 t of polymer and finish the outer pad in 1 hour. These figures seem like excessive hardware mass and speed, but they reflect the high program delay cost compared to the cheap transportation. In the future, high levels of commercial lunar activity may result in even higher program delay cost, because as the resource competition heats up and as real, commercial economic returns emerge, the cost of delay moves from being a political one to a very hard economic one.

*8.8 Cheap Transportation/Moderate Delay Cost*

The eighth case in Fig. 8(H) uses cheap transportation and moderate program delay cost. SiPo now has a total cost of only $65M (appropriated cost $30M). It uses 1.4 t of grading and compacting attachments to complete site preparation in 9 hours. It uses 13.1 t of sintering hardware including rovers to complete the inner pad in 44 hours. It uses 0.4 t of polymer application hardware (not including the rovers) to apply the 7.2 t of polymer to finish the outer pad in 4 hours.

*8.9 Cheap Transportation/No Delay Cost*

The ninth case in Fig. 8(I) uses cheap transportation and no program delay cost. SiPo has a total cost of $19M ($12M appropriated cost). It uses 239 kg of grading and compacting attachments to finish site preparation in 2.1 days. It uses 2.2 t of sintering hardware including rovers to complete the inner pad in 10.7 days. It uses 70 kg of polymer application hardware (not including the rovers) to finish the outer pad in 1 day.

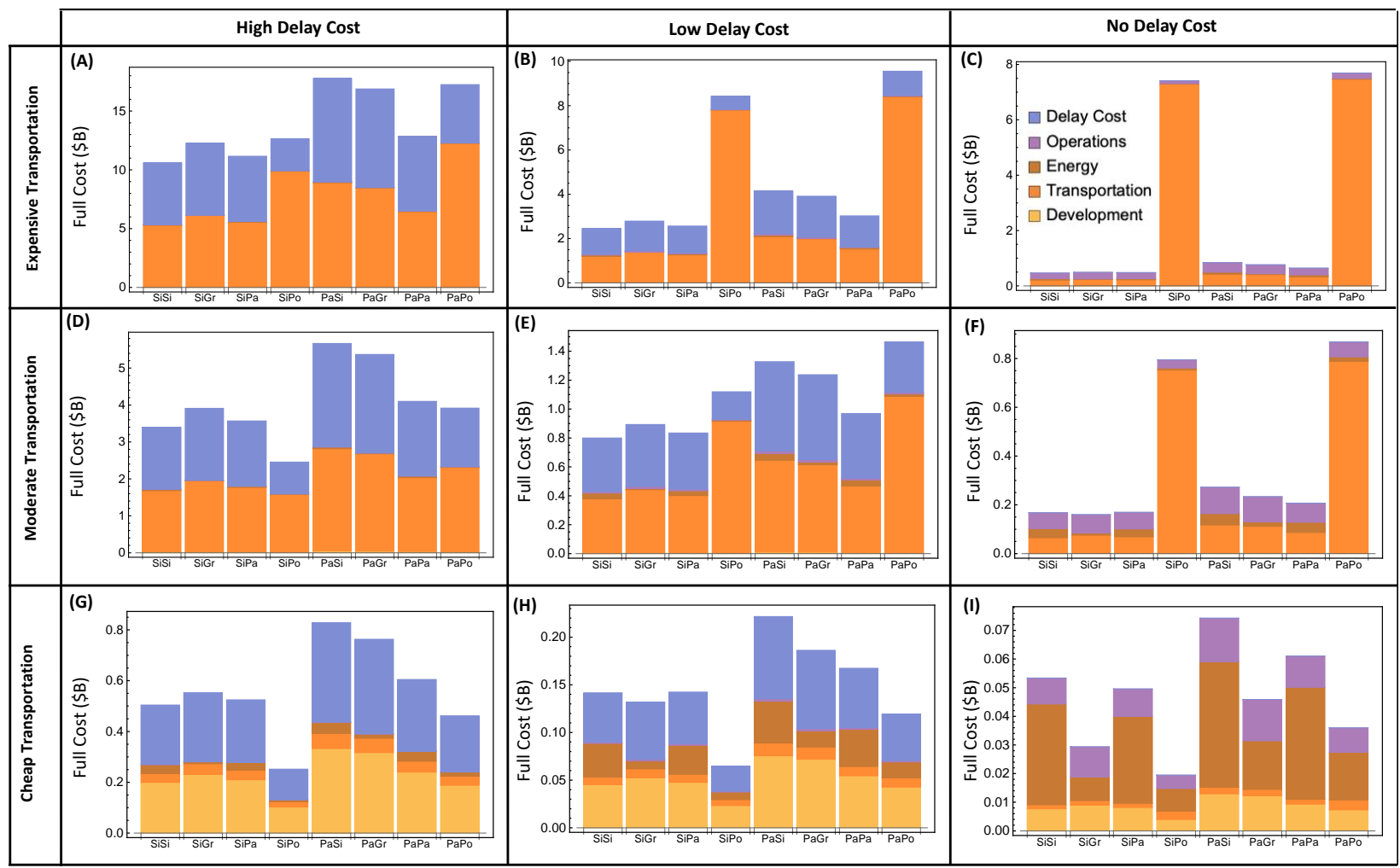

**Figure 8.** Optimized full cost of pad construction for pairs of construction methods in six economic scenarios. Note vertical axis scale is different on each graph. Expensive transportation is $1M/kg. Moderate transportation is $100K/kg. Cheap transportation is $300/kg. High program delay cost is for 25% activities delayed at a $3.2T-value lunar outpost. Moderate program delay cost is for 25% activities delayed at a $160B-value lunar outpost. No program delay cost means 0% of valued activities delayed (unrealistic).

## 9. Least Expensive Construction Method

In all cases, sintering the inner pad was favored over the use of pavers for the inner pad, although the uncertainties of the analysis could overturn this. Since they are close, choosing sintering will produce a valid estimate of the lowest cost. The choice for the outer pad depends on the cost of transportation. With cheap transportation cost, polymer infusion produces the least expensive outer pad. When transportation is expensive, then sintering is the least expensive for both inner and outer pads. Pavers and gravel/rock pads are both close to the cost of sintering for the outer pad. Figure 9 shows the total cost of optimized SiSi and optimized SiPo as a function of transportation cost for the baseline outpost that costs $160B with 4x expected value. SiPo becomes less expensive than SiSi when transportation cost drops below about ~$110K/kg. Figures 10 and 11 show the mass and construction time, respectively, corresponding to the cases in Fig. 9.

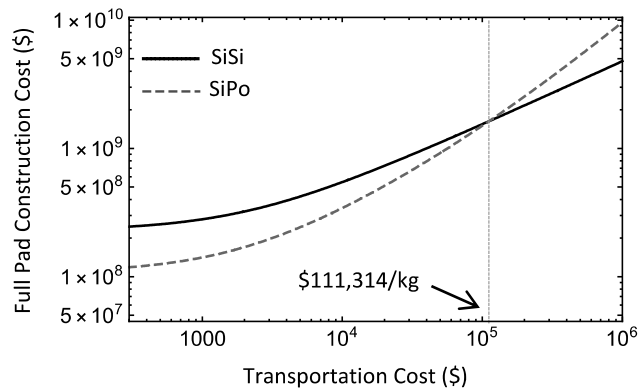

**Figure 9.** Cost of SiSi and SiPo for the baseline economic scenario but varying the transportation cost.

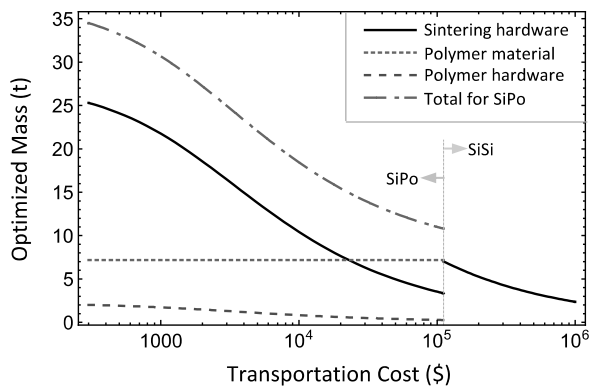

**Figure 10.** Mass of the least expensive system for the baseline economic scenario but varying the transportation cost.

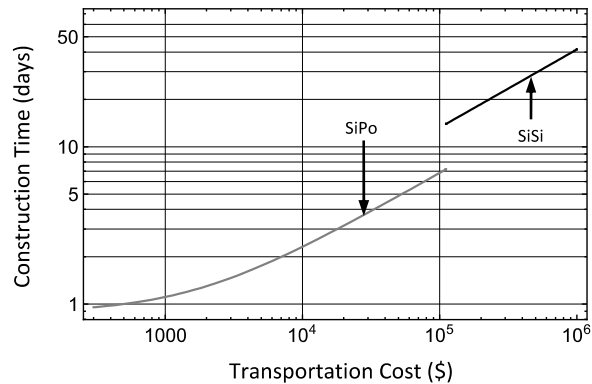

**Figure 11.** Construction time of the least expensive system for the baseline economic scenario but varying the transportation cost.

## 10. Examples of Cost-Optimized Construction Systems

Table 5 describes the optimized construction hardware for four specific transportation costs along the curves of Figs. 9–11. The optimized masses of hardware include the number of rovers needed to carry the construction attachments and the polymer material if applicable.

**Table 5. Example Optimized Construction Systems for Four Economic Scenarios**

| Item | Four Economic Scenarios with $160B Outpost, Expected 4x Value | | | |
|---|---|---|---|---|
| Transportation Cost | $1M/kg | $100K/kg | $10K/kg | $2K/kg |
| Method | SiSi | SiPo | SiPo | SiPo |
| Number of Rovers (~200 kg each) | 2 | 4 | 11 | 21 |
| Total Mass of Rover(s) | 492 kg | 833 kg | 2,311 kg | 4,116 kg |
| Sintering Hardware Mass | 1,641 kg | 2,778 kg | 7,705 kg | 13,721 kg |
| Polymer Application Systems Mass | 0 | 112 kg | 262 kg | 1,400 kg |
| Polymer Material Mass | 0 | 7,200 kg | 7,200 kg | 7,200 kg |
| Max Power Needed | 91.4 kW | 155 kW | 429 kW | 764 kW |
| Total Construction Time | 41.6 d | 6.8 d | 2.5 d | 1.4 d |
| Total Cost | $4,459M | $1,503M | $292M | $181M |
| Appropriated Cost | $2,383M | $1,110M | $143M | $81M |

## 11. Adjustments for Programmatic and Political Realism

The cases in Table 5 have very short construction times, large power requirements, large hardware masses, and large appropriated costs. These are the correct economic optimizations. However, the program delay cost and energy cost are unlikely to be as smoothly incrementable as assumed in this trade study method. Funding a large national space program is often as much of a political decision as an economic one. A lunar program may not be funded on a schedule that supports an optimal launch cadence that will adequately recover opportunity cost by accelerating tasks, and energy systems may be scaled according to programmatic decisions that cannot be responsive to each use of energy on the Moon. An alternative version of this trade study is presented in Table 6, which models this political realism using NASA's Artemis program.

The first three columns of Table 6 represent scenarios for Artemis with three transportation costs. The Artemis program plans to launch yearly, so we may assume landing pad construction must be completed in 270 days. This allows one month for unloading, checkout, and commissioning of the construction and solar power systems, and it reserves two months for contingency. This maximum allowable construction time is imposed as a constraint in the economic optimization. The maximum available power is also constrained to 50 kW in these examples. In this case, *appropriated* cost rather than the *total* cost is minimized. In the first two columns, the programmatic constraint on construction time becomes the limiting factor. In the third, the minimum appropriated cost is a balance of hardware mass and operations cost at a point that does not approach the programmatic constraints. We think this methodology produces more realistic estimates of the cost of lunar landing pads during the Artemis program, from $130M to $548M depending on transportation cost. It should be noted these estimates do not include the cost of technology maturation from the current TRL-3/4 to TRL-6. However, it is assumed that there are other applications for this technology and that some of that cost will be borne by other programs or via private sector investments.

**Table 6. Example Construction Systems for Five Scenarios with Program Constraints**

| Item | Five Scenarios with Programmatic Schedule and Power Constraints | | | | |
|---|---|---|---|---|---|
| Transportation Cost | $1M/kg | $300K/kg | $100K/kg | $10K/kg | $2K/kg |
| Schedule Constraint | ≤270 d | ≤270 d | ≤270 d | ≤120 d | ≤30 d |
| Power Constraint | ≤50 kW | ≤50 kW | ≤50 kW | ≤50 kW | ≤100 kW |
| Optimum Method | SiSi | SiSi | SiSi | SiSi | SiPo |
| Number of Rovers (~100 kg each) | 1 | 1 | 1 | 2 | 3 |
| Total Mass of Rover(s) | 100 kg | 100 kg | 134 kg | 269 kg | 356 kg |
| Sintering Hardware Mass | 316 kg | 316 kg | 445 kg | 897 kg | 1,710 kg |
| Polymer Application Systems Mass | – | – | – | – | 81 kg |
| Polymer Material Mass | – | – | – | – | 7,200 kg |
| Max Power Needed | 18 kW | 18 kW | 25 kW | 50 kW | 66 kW |
| Grading & Compacting Time | 11 d | 11 d | 8 d | 4 d | 3 d |
| Inner Sintering Time | 58 d | 58 d | 41 d | 20.5 d | 15.5 d |
| Outer Sintering Time | 201 d | 201 d | 142 d | 70.5 d | – |
| Outer Polymer Application Time | – | – | – | – | 1 d |
| Total Construction Time | 270 d | 270 d | 192 d | 95 d | 19.5 d |
| Appropriated Cost | $548M | $229M | $130M | $47.4M | $27.5M |

The fourth and fifth columns of Table 6 use the same method but with constraints that represent the developing cislunar economy. As transportation cost drops, the cost of doing everything in space will drop and the launch cadence will increase, so shorter construction times and larger power constraints are imposed. The results indicate that appropriated cost of constructing lunar landing pads may drop below $50M.

## 12. Strategic Technology Improvements

This trade study has also identified what improvements are strategic for each construction technique to become more competitive. These are listed in Table 7.

**Table 7. Technology Improvements for Competitiveness**

| Technique | | Most Strategic Improvement | |
|---|---|---|---|
| | | *If Expensive Transportation* | *If Cheap Transportation* |
| **Sintering** | *Inner* | (Currently best) | (Currently best) |
| | *Outer* | (Currently best) | Reduce mass of hardware per microwave power |
| **Pavers** | *Inner* | Increase construction speed per mass of hardware. Develop fast grouting method. | Lower energy usage |
| | *Outer* | Improve construction speed per mass of hardware | No delay cost: lower energy usage High delay cost: improve speed per mass of hardware |
| **Gravel** | *Inner* | Add a gas impermeable top layer; improve speed per mass of hardware | Add a gas impermeable top layer; improve speed per mass of hardware |
| | *Outer* | Improve construction speed per mass of hardware | Improve construction speed per mass of hardware |
| **Polymer** | *Inner* | Make capable of withstanding plume temperature; lower maintenance/repair cost; make the polymer in situ using lunar ice | Make capable of withstanding plume temperature; lower maintenance/repair cost |
| | *Outer* | Make the polymer in situ using lunar ice | (Currently best) |

## 13. Evaluation of the Trade Study Method

This method of performing a trade study in which all factors are converted into costs has produced insight that would not have been obtained from the standard trade study method. First, it demonstrated that reliability, energy, and operations costs are not significant cost-drivers in lunar landing pad construction. Without a direct comparison of quantified costs, their importance would possibly be exaggerated, even by experts. This could result in premature decisions to stop funding the more complicated technologies because of the perception that they are too complicated and thus would cost too much to maintain. That decision should be made only in context of the dominant overall costs (program delay cost and transportation cost). Second, this method provides a quantitative way to optimize the size-scaling of construction technologies, which is not provided by standard trade study methods. Third, it produced an estimate of the total and appropriated costs of lunar construction which can help inform early architectural decisions for the outpost and geopolitical decisions about projecting national presence onto the Moon and influencing international lunar agreements.

A weakness we found in this trade study method is the inherent assumption that all costs are smoothly incrementable, whereas recovery of program delay cost depends on the discrete nature of the launch cadence, and the real energy cost depends on high-level decisions about energy capacity of the overall outpost. These programmatic realities are usually beyond the scope of a technology trade study regardless the method that is used, but the economic method provides clear inputs to inform the programmatic decisions at least as well as the other methods, and arguably better. The second method (Table 6), where programmatic-level constraints for schedule and power were imposed while minimizing the appropriated cost instead of total cost, was found to be an improvement on the basic method in the programmatically constrained cases, and it can be used together with the basic method to inform decisions.

## 14. Limitations of the Study and Future Work

The construction methods studied here are not exhaustive nor do they provide the final word on the best construction method. The parameters used in this tool are based on comparisons to terrestrial technologies; their values can be improved after the space construction technologies have been prototyped and measured experimentally. The initial version of this trade study was developed in support of the Robotic Lunar Surface Operations 2 (RLSO2) study [69,70]. This new work has extended the methodology of that study and programmed it into a Mathematica notebook, which can be made available to qualified aerospace companies and researchers to support lunar development. Technology developers can use this model with improved parameters to evaluate their technologies or may develop performance targets for new construction methods to ensure they are competitive.

## 15. Conclusions

Finding the minimum cost to build a lunar landing pad depends on optimizing the mass of the construction systems to balance the transportation cost with the program delay cost. In the context of a lunar surface outpost with $100K/kg to the lunar surface, which many expect to occur during NASA's Artemis program, then a landing pad will require only about $299M on the budget line item, which is the cost of a NASA Discovery Program mission. Considering the significance of building the infrastructure of civilization on another planet, this level of cost is easily justified. Some space companies and analyses are projecting much lower transportation costs in the next decade, perhaps reaching $10K/kg or lower. If so, then a landing pad at an outpost may require only $46M or less on the program budget. Among the construction methods considered here, microwave sintering is most economical for both the inner and outer zones of the landing pad. However, when transportation costs drop below about $110K/kg, then a hybrid method becomes more economical with sintering for the inner zone and polymer infusion for the outer zone. Several construction techniques are close enough in cost that they are within the range of uncertainty to be competitive with sintering and polymer infusion. New space-capable nuclear power systems are under development and the availability of nuclear power on the lunar surface might significantly change the cost of power and remove the complexities of the solar cycle from the construction process. Further innovations may still be decisive in determining the best construction method, so it is prudent to continue investing in a variety of techniques at this time.

**Declaration of Competing Interest**

The University of Central Florida has applied for patent of a technology invented by Philip Metzger to reduce the energy of microwave sintering lunar soil. Metzger is partially supported by NASA Small Business Technology Transfer (STTR) program award number T7.04-2630 (STTR 2021-1)), which is developing that technology. Microwave sintering in general is discussed in this manuscript, but the improvements expected from the new technology are not included or assessed here.

**Acknowledgements**

The authors wish to thank Jim Bell of the Arizona State University, Tory Bruno of the United Launch Alliance, Lindy Elkins-Tanton of the Arizona State University, Lars Hoffman of Rocket Lab Global Launch Services, Scott Hubbard of Stanford University, Rob Mueller of NASA Kennedy Space Center, and Steve Squyres of Blue Origin for helpful discussions.

**Funding**

This work was supported in part by NASA Solar System Exploration Virtual Institute cooperative agreement award NNA14AB05A, "Center for Lunar and Asteroid Surface Science," and by NASA Small Business Technology Transfer (STTR) program award number T7.04-2630 (STTR 2021-1), "High Efficiency Sintering via Beneficiation of the Building Material."

## Appendix A: Model Parameters and Calculation Methods

The parameters used in this study were developed to represent best estimates of the current state of the art. They are listed in Table A-1. To enable evaluation of other cases, to evaluate progress in the technologies, and to test the sensitivity of these parameter values, the model will be made available to qualified aerospace companies and the researchers.

**Table A-1. Trade Study Input Parameters**

| Parameter | Value | Units |
|---|---|---|
| *Basic Data* | | |
| Pad Inner Zone Radius | 12 | m |
| Pad Outer Zone Radius | 27 | m |
| Density of Soil Before Compaction | 1500 | kg/m$^3$ |
| Specific Gravity of Minerals in Regolith | 3100 | kg/m$^3$ |
| *Roving, Grading, and Compacting* | | |
| Rover Mass | 300 | kg |
| Driving Speed | 1 | m/s |
| Roving Specific Energy per Distance | 2.5 | J/kg/m |
| Grading Rate | 0.1 | m$^2$/s |
| Compacting Rate | 0.05 | m$^2$/s |
| Grader Blade Width | 1.5 | m |
| Grading Energy | 0.0167 | kWh/m |
| Compactor Mass | 200 | kg |
| Grading Blade Mass | 300 | kg |
| Density of Regolith After Compaction | 2200 | kg/m$^3$ |
| *Gravel/Rock Pad (Breakwater)* | | |
| Thickness of Rock Pad | 22.86 | cm |
| Number of Rock Layers | 4 | – |
| Mass of Raking/Sorting Rover | 1000 | kg |
| Mass of Rock Laying Rover | 600 | kg |
| Fraction of regolith that is usable rock | 4 | wt% |
| Bulk density of packed rock pad | 2200 | kg |
| Rock rake width | 1 | m |
| Rock raking depth | 0.1016 | m |
| Rock raking speed | 0.667 | m/s |
| Time to lay rock | 300 | s/m$^2$ |
| Width of rock laying device | 1 | m |
| Power of rock rake at deepest depth | 982 | W |
| Rock sorter (trommel) energy | 0.433 | kWh/t |
| Rock load per trip | 1000 | kg |
| *Microwave Sintering* | | |
| Inner Pad Sintered Thickness | 7.62 | cm |
| Outer Pad Sintered Thickness | 2.54 | cm |
| Maximum Payload (Sintering Equipment) Per Rover | 1000 | kg |
| Density of Sintered Pad | 2200 | kg/m$^3$ |
| Available Power for Sintering (outpost capability) | 200 | kW |
| Magnetron Efficiency | 0.5 | – |
| Terrestrial Magnetron Power/Mass Ratio | 3/260 | kW/kg |
| Spaceflight Optimization Factor for Magnetron Mass | 0.2 | – |
| Inner Pad Energy Application | 21.73 | kWh/m$^2$ |
| Outer Pad Energy Application | 18.45 | kWh/m$^2$ |
| *Polymer Infusion* | | |
| Inner Polymer Pad Thickness | 5.08 | cm |
| Outer Polymer Pad Thickness | 2.54 | cm |
| Outer Polymer Mass Fraction | 7 | wt% |
| Inner Polymer Mass Fraction | 11.66 | wt% |
| Mass of Sprayer/Infusion Assembly | 100 | kg |
| Mass of Polymer in Full Rover Tank | 1000 | kg |
| Rover Tank Refill Time | 30 | minutes |
| Polymer Application Time | 10 | s/m$^2$ |
| Spray Width | 1 | m |
| Polymer Density | 1000 | kg/m$^3$ |
| Distance Polymer Storage to Pad | 1 | km |
| *Pavers* | | |
| Inner Pad Paver Thickness | 7.62 | cm |
| Outer Pad Paver Thickness | 2.54 | cm |

| Paver Horizontal Dimension (square shape) | 45.72 | cm |
| Paver Material Density | 2200 | kg/m$^3$ |
| Oven and Associated Mechanisms Mass | 1000 | kg |
| Oven Distance to Pad | 20 | m |
| Paver Installation Robotic Arm (on rover) | 100 | kg |
| Feedstock Excavator Implement Mass (on rover) | 100 | kg |
| Excavator Digging Bucket Width | 0.5 | m |
| Excavator Digging Bucket Depth (Bite Depth) | 0.3 | m |
| Feedstock Excavation Rate | 2.286 | kg/s |
| Feedstock Excavator Power | 4 | kW |
| Feedstock Load on Rover Per Trip | 1000 | kg |
| Time to Fill Molds and Place into Oven Per Paver | 30 | s |
| Oven Sintering Temperature | 1120 | °C |
| Oven Starting Temperature | 100 | °C |
| Average Thermal Conductivity in Packed Molds | 346.99 | mW/m/K |
| Average Specific Heat in Packed Molds | 1095.19 | J/kg/K |
| Oven Cooling Time Factor | 0.5 | – |
| Oven Energy Efficiency | 0.6 | – |
| Transfer Time from Oven to Rover Per Paver | 15 | s |
| Max Load of Pavers on Rover When Hauling | 1000 | kg |
| Installation Time Per Paver | 60 | s |
| Robot Power to Install Pavers | 400 | W |
| Grout Density | 1500 | kg/m$^3$ |
| Grout Bead Radius | 3 | mm |
| Grout Insertion Rate | 1 | cm/s |

*Roving, Grading, and Compacting*

The power needed to compact the soil per square meter was based upon a commercially available compactor [71]. The time required to compact the soil was estimated based on experience of one of the authors (Metzger) working as a regolith judge in large arenas of lunar soil simulant in NASA's Lunabotics mining competition [72]. The time-averaged power for grading was chosen to match an electric grading robot [73].

Roving energy per kilogram mass per distance was based on the 20 kg packbot, rounding up values from Table 3.5 by Broderick [74]. Roving energy is calculated by multiplying the energy per kilogram per distance by the mass of the rover with its implements and/or mass of materials and by the distance traveled.

Excavator blade force was based on Gallo et al. [75] with 6 cm depth of blade. This was divided by 6 assuming the forces scale with gravity. This would not be correct for small-scale digging where cohesion dominates, but for large-scale digging where mass forces dominate this is a good approximation. Blade energy is blade force times distance. Grading energy is blade energy plus roving energy including the mass of the rover and the excavator blade.

Driving energy during the compaction operations is roving energy including the mass of the rover and the compactor device. Driving distance during grading or compacting is the area of the pad divided by the width of the grader blade (assumed the same width as the compaction device).

Compaction time and grading time are the compaction or grading rate divided into the area of the pad. Compaction energy is driving energy with the mass of the compactor over the driving distance plus the compactor power times the compaction time. Grading energy is the driving energy with the mass of the grading blade over the driving distance. Grading or compacting power requirements are the grading energy or compacting energy divided by grading time or compacting time.

*Gravel/Rock Pad*

The rock raking energy is estimated by analogy to soil tillage machines. Rock raking speed was based on a commercial power tiller [76]. Rock raking energy was estimated upon the basis of commercial power harrows scaled from terrestrial farm soil to lunar soil using the Balovnev [77] bucket force equation scaled to the size of a raking tine. Wilkinson and DeGennaro [78] applied this equation to lunar soil simulant and validated it against experimental data for a flat plate pushing the simulant. Here the plate is defined to have 0.635 cm width (approximately the width of a single tine from a power harrow) with vertical orientation. To calibrate the terrestrial comparison, gravity was set to 9.81 m/s² and the soil parameters to 29.59 deg internal friction angle and 42.1 Pa cohesion, which are the averages for farmland soil from Li et al [79], and 1340 kg/m³ for the bulk density as an average for pre-tilled farmland [80]. For the lunar comparison the equation was parameterized for lunar soil and lunar gravity 1.622 m/s² with the same rake tine. The bulk density of lunar soil varies dramatically over the raking depth, so a correlation from Apollo soil data [80] was used for the bulk density versus depth:

$$\rho(z) = 1920 \frac{z + 0.122}{z + 0.18} \text{ kg/m}^3$$

(A-1)

where depth $z$ is in meters. From this the relative density is,

$$D_R(z) = \left(\frac{z + 0.122}{z + 0.18} - \frac{0.122}{0.18}\right) \Big/ \left(1 - \frac{0.122}{0.18}\right) \times 100\%$$

(A-2)

Data from the Lunar Sourcebook [81] measured in a basaltic lunar soil simulant are approximately replicated below in Fig. A-1. This provides the friction angle and cohesion to use in the Balovnev equation as a function of the relative density and thus as a function of the depth.

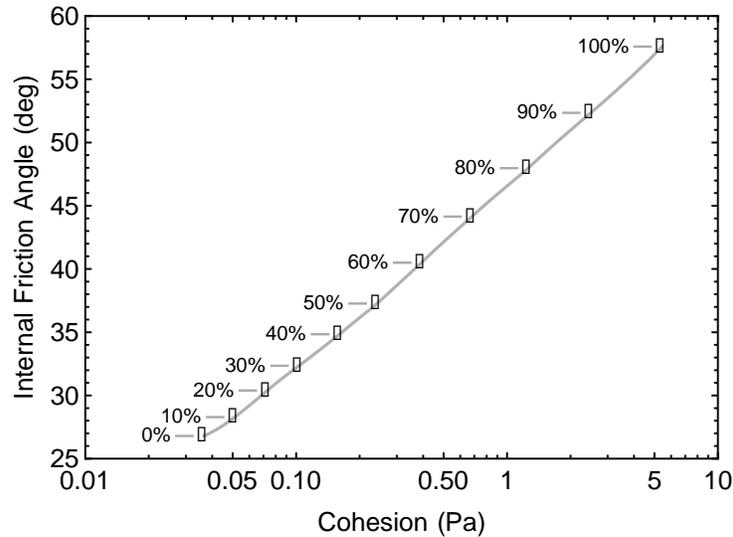

**Figure A-1.** Parametric plot of friction and cohesion as functions of relative density (annotated circles on plot) for a lunar simulant, following Heiken et al. [81] which follows Mitchell et al. [82,83].

With these parameters, the force on the rake tine in lunar soil was divided by the force on the same rake tine in terrestrial farm soil, and the ratio is plotted verses depth of the tine into the soil in Fig. A-2. This ratio is fairly constant and averages to 0.199 across the range of interest. This ratio is close to the ratio of gravities, 0.165, confirming the *a priori* belief that gravity is dominant in the scaling although soil type also has an influence. With equal raking speed in both terrestrial and lunar cases, the energy of raking will scale according to this reduction in force. This methodology is adequate since energy turns out to be less important than other factors, so the degree of inaccuracy does not affect the outcome of this trade.

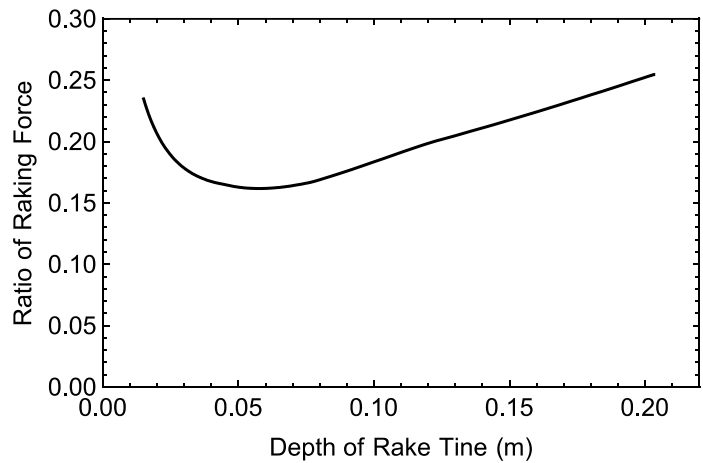

**Figure A-2**. Ratio of raking force in lunar soil to terrestrial farm soil versus depth or rake in the soil, calculated by the Balovenev equation.

The reference commercial harrow [84] uses about 80 hp (60 kW), has a full raking width of 3 m, and has tines that penetrate 20.32 cm (8 inches) deep into the soil. The tine motions are complicated but were designed to minimize energy while fully disaggregating the soil over that width, so the operating width (not the number of tines) will be used to scale the total raking power,

$$P_{\text{raking}} = (\text{Rake Factor}) \cdot (\text{Rake Width}) \cdot (\text{Force Per Tine}) \cdot (\text{Raking Speed})$$

(A-3)

From this we can solve for the Rake Factor = ~840/m, which accounts for machine inefficiencies and the number of tines per meter. The lunar case is calculated using the same Rake Factor and Raking Speed but with its own Rake Width (assumed 1 m) and its own Force Per Tine as a function of depth in the lunar soil calculated by the Balovnev equation using lunar soil parameters. For the deepest raking described below, the lunar raking power turns out to be equal to the harrow's power divided by 26, so raking in the fluffy upper layers of lunar soil in lunar gravity turns out to provide a significant energy reduction.

The largest rock size that is needed is 15 cm diameter and the smallest is 1.5 cm (Paul van Susante, personal communication; see also [36]). The model here assumes four rock sizes distributed logarithmically across that range, 1.5 cm, 3.23 cm, 6.96 cm, and 15 cm. Generally, rocks are at least partially embedded in the soil so larger rocks require deeper raking to dislodge them, requiring more energy. Different sized rocks must be collected for the rock landing pad, and more area must be raked to obtain the smaller rocks than the larger rocks (see below), so there is no need to rake at the greatest depth over the entire area. After enough larger rocks have been collected, the raking will continue at successively shallower depths. Regardless of the raking depth, some rocks will be encountered that cannot be removed without greater penetration, so a raking operation must always be prepared to retract the rake and skip over such rocks. This strategy will enable the successively shallower raking that minimizes energy. The raking forces for the four depths to extract these size rocks using the Balovnev equation are 8.3, 24.7, 104, and 667 N.

Paul van Susante (personal communication) provided areas that must be raked to collect enough of each size rock. Fitting to his data finds a power law to predict the raking area for each rock size,

$$A_{\text{raking}} = 37{,}772\, D^{-0.535}\ (\text{m}^2)$$

(A-4)

where rock diameter $D$ is in meters. These areas divided by the width of the rake provide the raking path length for each depth (to collect enough of each rock size). The energies to rake over each of those four path lengths are summed to determine the total raking energy. The amounts of time to rake over those distances at the raking speed are summed to determine the total raking time.

The rock sorter energy is based on a commercial trommel [85]. Sorting time occurs simultaneously with raking time, but for the purposes of subsystem reliability calculations we assume sorting requires effectively half as much operating time as raking.

Energy to haul rock and perform raking and rock laying calculations is similar to the grading and compacting calculations, except the loads are based on the amount of rock being hauled, which increases linearly during raking operations and decreases linearly while the rock is being deposited.

*Microwave Sintering*

The mass of the microwave system was estimated by two types of commercial microwave ovens. Lab sintering ovens at 260 kg produce 3 kW power, dividing the mass by two to remove the lab chassis predicts 23.08 W/kg. Cooking ovens at 45.5 kg (100 lbm) can produce 1 kW, which predicts 22.0 W/kg, essentially the same as the other estimate. The mass of the system is then multiplied by a Space Mass Optimization factor = 0.2, representing spacecraft design reducing the mass of a terrestrial system since GaN FETs used for lunar soil sintering have obtained this mass reduction (Dennis Wingo, personal communication).

The number of rovers needed to perform the sintering is calculated by dividing the mass of sintering equipment by the nominal payload mass that an individual rover is expected to carry, then rounding.

The roving energy as the rovers carry the mass of sintering equipment over the required distance to sinter the pad is negligible compared to the microwaving energy itself (five orders of magnitude) so it is neglected.

It is important for this study to have a reasonable estimate of the energy of microwave sintering. This was developed by writing a physics-based finite difference model of microwave absorption. It simulates the passage of microwave energy from the top to the bottom of a soil column as a plane wave. The temperature at each point in the soil column determines microwave absorption rate at that depth and thus the microwave energy flux that transmits through to the next depth. In each time step, the absorbed energy at a location and the specific heat at the current temperature at that location determine how much the temperature increases by the next time step. Lunar soil is a good insulator [86] and microwave heating is relatively fast compared to the rate of a thermal wave propagation in lunar soil, so thermal conductivity is neglected in this approximation, which is adequate for a trade study. In future work, the model will be improved to include the physics of thermal conductivity.

The sintering experiments by Allan et al. [50] measured the microwave loss tangent and half-power depth in lunar soil simulant as a function of the soil's temperature. The data only went to 1079 °C, but sintering requires raising the temperature to about 1200 °C. The data show the half-power depth is small and not changing rapidly by the time the soil reaches 1079 °C, so we assume it stays approximately unchanged from 1079 °C to 1200 °C. The half-power depth was converted to an exponential decay constant as shown in Fig. 2. For the heat capacity of lunar soil, available data sets with actual lunar soil do not go nearly high enough to simulate sintering physics. The specific heat of basalt approximates that of lunar soil, so this study uses the data and the empirical fitting function of basalt from Bouhifd et al. [87]. The fitting function is

$$C(T) = 2337 - 0.2773\,T + \frac{220.2 \times 10^5}{T^2} - \frac{29{,}760}{\sqrt{T}}$$

(A-5)

where the specific heat $C$ is in J/kg/°C and $T$ is in °C. This is shown in Fig. A-3.

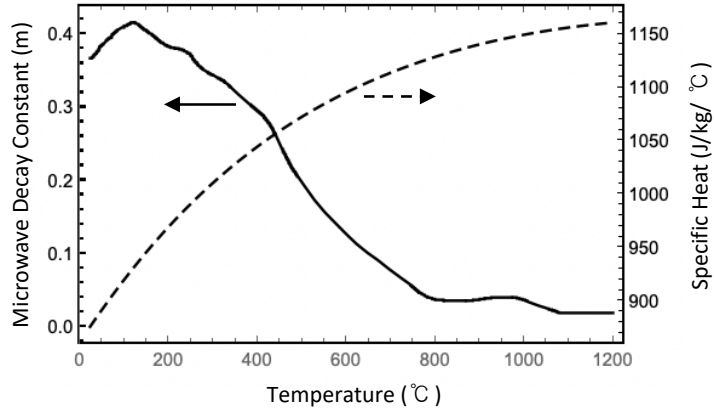

**Figure A-3.** Microwave decay constant and specific heat used to model microwave sintering of lunar soil.

The density of lunar soil varies in the soil column, but for landing pad construction the soil will be graded and compacted prior to sintering so the model assumes a constant bulk density of 2200 kg/m³ in the sintering zone. The starting temperature of the lunar soil is assumed to be 127 °C. The incident microwave energy flux is assumed to be 200 kW/m² for a baseline case, which is achievable using a magnetron and a horn antenna. (This flux was greatly reduced in most optimized versions of the construction system, see main text.) The model uses finite difference cells of 1 mm thickness and timesteps of 0.3 s. It was time-stepped until the average temperature within the desired sintering depth reached 1200 °C and we assume that thermal conductivity (which is much higher in this zone where the temperature is higher) will tend to average out the temperature in this zone. This approximation is deemed adequate for this early trade study, and it will be improved in future work. Depths of 7.62 cm and 2.54 cm were used for the inner and outer pads, respectively. An example of the resulting temperature profile after microwave application is shown in Fig. A-4. This model found that an energy per area of 21.73 kWh/m² was needed to produce a 3 cm thick sintered layer, or 18.45 kWh/m² for a 1 cm thick sinter, indicating that not much energy will be saved by keeping the outer pad thinner. Nevertheless, we use the smaller value for the outer pad.

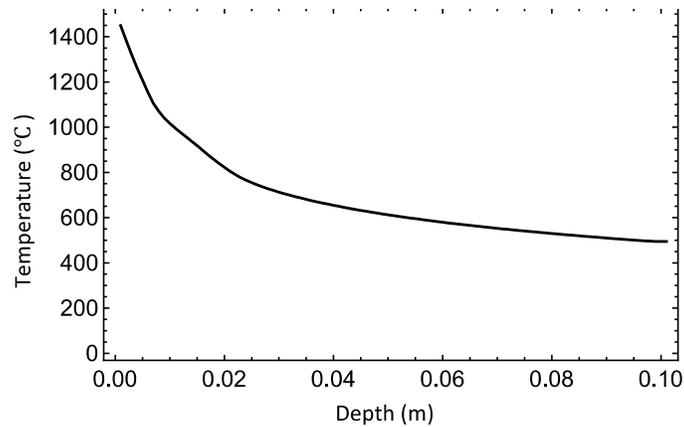

**Figure A-4.** Model's predicted temperature profile after microwaving 200 kW/m² at 2.45 GHz into lunar soil for 5.34 minutes. The average temperature is above the sintering temperature in the top 1 cm, but we assume thermal conductivity (otherwise neglected in this model) averages the temperatures over that length scale.

*Polymer Infusion*

For polymer in the inner pad, the fraction of mass of polymer is based on a calculation where the polymer completely fills the pore space of the compacted soil. For polymer in the outer pad, the mass fraction of polymer was based roughly upon similar cases in the heat shield project by Hogue et al. [44]. The energy of polymer infusion includes the amount of roving to refill the tank the number of times that are necessary while carrying the mass of polymer one-way and with decreasing polymer load during the infusion into the pad.

*Pavers*

The feedstock excavating time is calculated as the volume of soil that must be excavated divided by the excavation rate. The power and rate for excavating feedstock for the paver maker was based upon the 4 kW electric excavator studied by Nevrly et al. [73] roughly estimating it can excavate 16 kg in 7 seconds based upon digging bucket size and typical excavation kinematics. The excavating energy is the excavating power times the excavating time.

Excavation depth is assumed to be 30 cm because the typical lunar soil density below 30 cm makes it more difficult to excavate. The excavation area is the total volume of soil that must be excavated divided by the excavation depth. The excavating path length is the excavation area divided by the width of the digging bucket. Excavation is assumed to take place adjacent to the oven and takes place in a square area, so the average hauling distance of excavated feedstock is the average of the Pythagorean equation over that square to the midpoint of one of its sides, which comes out to 0.593 times the square root of the excavation area. This is the order of magnitude of only about 10 meters. The average hauling distance of pavers to the pad is the average of the Pythagorean equation over the area of the pad to the oven. If it is assumed the oven is 20 meters from the edge of the landing pad to facilitate construction, that comes out to the order of magnitude of 50 meters average hauling distance of pavers for a 27-meter pad. It may be safe to put the oven so close to the pad because it can be moved prior to operational use of the pad, or it can be left in place since the pad will prevent blowing ejecta and the oven should thus be safe.

The number of excavating cycles (driving to/from the excavation site and returning with feedstock for the oven molds) is based on the mass of all the pavers divided by the payload mass of the rover. This assumes surge control in the form of a hopper and regolith feed system to move soil from the hopper into the oven molds. The mass of the hopper and feed system is included in the oven mass.

The total feedstock hauling time is the average distance for excavating to the oven multiplied by the number of excavating cycles divided by the driving speed. The total paver hauling energy is calculated similarly. The total hauling energy is calculated similarly to the prior cases. The total installation time is the total paver hauling time plus the number of pavers divided by the installation rate.

The average specific heat $\langle C \rangle$ of the regolith for the paver calculations was determined by averaging the specific heat over the range of temperatures from the starting temperature of 127 °C to the ending temperature of 1200 °C using the curve fit of Bouhifd et al. [87] for basalt.

The average thermal conductivity $\langle \kappa \rangle$ over this temperature range is roughly approximated as 2/3 value at the high temperature and 1/3 the value at the low temperature from Eq. 14 of Metzger, et al. [86].

The total baking energy is the mass of the pavers times the average specific heat divided by the oven efficiency. The oven efficiency is based on typical, large-scale, terrestrial brick-making ovens surveyed by da Graca Carvalho and Nogueira [88].

The time to bake the pavers is determined as the time it takes the center of a paver to reach sintering temperature, assuming heat conducts in from both top and bottom of the paver. This is calculated from the heat diffusion equation as

$$t_{\text{heating}} = 6 \left[ \left(\frac{T}{2}\right)^2 \frac{\rho \langle C \rangle}{2 \langle \kappa \rangle} \right]$$

(A-6)

where $T$ is paver thickness, $\rho$ is the compacted bulk density of the soil in the paver mold, and the quantity in the square brackets is the exponential time constant. Six exponential time constants are allowed to reach oven temperature in the center of the paver. The oven cooling time before removing the pavers is approximated as a constant factor times the heating time.

Excavating and hauling can be done in parallel with baking except for the first oven batch, which cannot start until the requisite amount of excavation and hauling has delivered its feedstock, and except for the final paver installation, which cannot start until after the last oven batch has completed. The total pad construction time is therefore the sum of the long-pole process with the one batch that straddles the beginning and the end,

$$t = \frac{(N-1)}{N}\text{Max}(t_E + t_{FH}, t_B, t_G + t_C + t_{PHI}) + \frac{1}{N}(t_E + t_{FH} + t_B + t_{PHI})$$

(A-7)

where $N$ = the number of oven batches, $t_E$ = total excavation time, $t_{FH}$ = total feedstock hauling time, $t_B$ = oven baking and cooling time (total of all batches), $t_G + t_C$ = grading plus compaction time, and $t_{PHI}$ = paver hauling and installation time.

The power for the robotic arm on an excavator to install pavers was based on a study of electrical power needs by various types of robots Barnett et al. [89].

The grout density is based on terrestrial grout [90]. The grout bead radius is a rough estimate of the necessary volume of material to fill paver crevices. Grout mass is calculated as circumference of a paver multiplied by the number of pavers divided by two (since shared edges are grouted only once) multiplied by bead cross section and density. Grout application rate is a rough estimate based on experience extruding pastes. Grout application power is assumed to be the same as paver laying power.

## **Appendix B: Reliability Calculations**

The reliability calculation follows the Feasibility of Objective Technique of MIL-HDBK-338B [63]. Engineering experience is used to apply a rating from 1 to 10 in columns A, B, and D of Table B-1. The method was modified for column C because actual operating time predictions are available from the trade study model, and they were dominated by the baking operation (rated 10) so most other operations were rated 1, losing numerical distinction between them. This was improved by using non-integer values 0 to 10, which were scaled linearly from the predicted operating times. Column E is the relative failure rate, which is calculated as the product of Columns A through D. The operating times depend on the scaling of the construction systems, so the reliability calculations are specific to a particular optimization. The calculations shown here are for the Baseline Economic Scenario. Ideally, these calculations would be iterated with the optimization for each scenario because reliability cost will affect the optimization and vice versa. However, as we show here the reliability has negligible impact on all but the most extreme case of low transportation cost, and that will be farther in the future when landing pad construction methods have become much more reliable. Therefore, we show that reliability does not play a significant role in the trade and iteration is not necessary.

The results in column E of Table B-1 were used to populate column A of Table B-2. The subsystems in Table B-1 (individual rows) were added together as appropriate to constitute the various construction systems. For example, grading, compacting, and hauling were aggregated into Rover Operations, since the rover platform is the key element in them all. Each case of Rover Operations in Table B-2 may have a different failure rate than the other cases because the time of performance for the rover is different for each construction method, as reflected in Table B-1. To calculate column B in Table B-2, the failure rate in each row of Column A was divided by the sum of the rows for that entire construction method. (Column B will be needed to decide how much improvement is needed in each subsystem of a construction method to reach the target reliability at the minimum cost.) Column C was calculated by the following equation,

$$R_{i,0} = e^{-\lambda_{\text{rel},i}/\Lambda}$$

(B-1)

where $R_{i,0}$ = the reliability of the $i$th subsystem if it were made using standard components and design resilience, $\lambda_{\text{rel},i}$ = the relative failure rate of the $i$th subsystem given in column A, and $\Lambda$ is a normalization constant. Following MIL-HDBK-338B, the value of $\Lambda$ is chosen such that it produces the expected baseline reliability. Here, the relative reliable is highest for SiPo, so $\Lambda$ = 4337.9 is chosen such that SiPo overall reliability is 99.0%, which we presume the program specifies as the landing pad requirement. This in effect defines the baseline quality of components and design resilience needed to achieve 99%. Next, the calculation follows the Minimization of Effort Algorithm in MIL-HDBK-338B to determine the target reliability that must be achieved in each subsystem such that the overall construction method achieves 99% at minimum cost (column D). Finally, the method determines the additional cost needed to achieve this for each subsystem which are combined in weighted sum to determine the additional cost to the overall system (column E). These last two steps were described in the main text at Eq. 2.

**Table B-1. Cost of Reliability Calculations, Part 1.**

| Subsystem | A. Intricacy Rating (1-10) | B. State-of-the-Art Rating (1-10) | Operating Hours | C. Performance Time Rating (0-10) | D. Environment Rating (1-10) | E. Relative Failure Rate (events per unspecified time) |
|---|---|---|---|---|---|---|
| Grading & Compacting Inner Zone | 4 | 3 | 3.8 | 0.06 | 4 | 3.01 |
| Grading & Compacting Outer Zone | 4 | 3 | 15.3 | 0.25 | 4 | 12.11 |
| Sintering Inner Zone | 3 | 5 | 22.42 | 0.37 | 5 | 27.72 |
| Sintering Outer Zone | 3 | 5 | 41.51 | 0.68 | 5 | 51.32 |
| Sintering Both Zones | 3 | 5 | 47.18 | 0.78 | 5 | 58.33 |
| Polymer Outer Zone | 3 | 2 | 2.59 | 0.04 | 3 | 0.77 |
| *Pavers Inner Zone Subsystems:* | | | | | | |
| Excavating | 6 | 5 | 10.07 | 0.17 | 10 | 49.78 |
| Hauling Reg & Pavers | 6 | 2 | 2.41 | 0.04 | 6 | 2.86 |
| Oven Robotics | 10 | 8 | 29.98 | 0.49 | 4 | 158.15 |
| Baking | 1 | 1 | 111.82 | 1.84 | 2 | 3.69 |
| Laying Pavers | 6 | 4 | 39.50 | 0.65 | 3 | 46.88 |
| Grouting | 6 | 10 | 60.17 | 0.99 | 2 | 119.04 |
| *Pavers Outer Zone Subsystems:* | | | | | | |
| Excavating | 6 | 5 | 18.42 | 0.30 | 10 | 91.11 |
| Hauling Reg & Pavers | 6 | 2 | 4.86 | 0.08 | 6 | 5.77 |
| Oven Robotics | 10 | 8 | 162.40 | 2.68 | 4 | 856.75 |
| Baking | 1 | 1 | 606.58 | 10.00 | 2 | 20.00 |
| Laying Pavers | 6 | 4 | 216.05 | 3.56 | 3 | 256.44 |
| *Pavers Both Zones Subsystems:* | | | | | | |
| Excavating | 6 | 5 | 19.06 | 0.31 | 10 | 94.28 |
| Hauling Reg & Pavers | 6 | 2 | 4.83 | 0.08 | 6 | 5.73 |
| Oven Robotics | 10 | 8 | 120.88 | 1.99 | 4 | 637.68 |
| Baking | 1 | 1 | 451.36 | 7.44 | 2 | 14.88 |
| Laying Pavers | 6 | 4 | 160.50 | 2.65 | 3 | 190.51 |
| Grouting | 6 | 10 | 48.32 | 0.80 | 2 | 95.58 |

| Gravel Outer Zone Subsystems: | | | | | | |
|---|---|---|---|---|---|---|
| Raking | 6 | 5 | 6.75 | 0.11 | 10 | 33.40 |
| Sorting | 5 | 5 | 3.38 | 0.06 | 8 | 11.13 |
| Laying rock | 6 | 4 | 27.78 | 0.46 | 6 | 65.94 |
| Hauling | 6 | 2 | 4.03 | 0.07 | 6 | 4.78 |

Table B-2. Cost of Reliability Calculations, Part 2.

| System or Subsystem | A. Relative Failures (some rows are combinations of several subsystems from Table B-1) | B. Fraction of failures (within each construction technique) due to each subsystem | C. Subsystem or System Reliability Prior to Improvement | D. Goals for Subsystem Reliability Improvement | E. Subsystem (or System) Cost Function to Meet Reliability Goals |
|---|---|---|---|---|---|
| **SiSi** | | | | | |
| Rover operations[1] | 15.11 | 16.05% | 99.65% | 99.65% | 1.000 |
| Sintering | 79.04 | 83.95% | 98.19% | 99.35% | 1.375 |
| ***OVERALL SYSTEM*** | ***94.15*** | | ***97.85%*** | | ***1.305*** |
| **SiPa** | | | | | |
| Rover operations[2] | 20.89 | 1.64% | 99.52% | 99.83% | 1.385 |
| Sintering | 27.72 | 2.18% | 99.36% | 99.83% | 1.446 |
| Excavating | 91.11 | 7.16% | 97.92% | 99.83% | 1.584 |
| Oven robotics | 856.75 | 67.31% | 82.08% | 99.83% | 1.641 |
| Baking | 20.00 | 1.57% | 99.54% | 99.83% | 1.374 |
| Laying pavers | 256.44 | 20.15% | 94.26% | 99.83% | 1.625 |
| ***OVERALL SYSTEM*** | ***1272.91*** | | ***74.57%*** | | ***1.461*** |
| **SiPo** | | | | | |
| Rover operations[1] | 15.11 | 34.67% | 99.65% | 99.65% | 1.000 |
| Sintering | 27.72 | 63.57% | 99.36% | 99.36% | 1.000 |
| Polymer application | 0.77 | 1.76% | 99.98% | 99.98% | 1.000 |
| ***OVERALL SYSTEM*** | ***43.60*** | | ***99.00%*** | | ***1.000*** |
| **SiGr** | | | | | |
| Rover operations[3] | 19.90 | 12.59% | 99.54% | 99.75% | 1.253 |
| Sintering | 27.72 | 17.53% | 99.36% | 99.75% | 1.354 |
| Raking and sorting rocks | 44.54 | 28.17% | 98.98% | 99.75% | 1.458 |
| Laying gravel/rock | 65.94 | 41.71% | 98.49% | 99.75% | 1.517 |
| ***OVERALL SYSTEM*** | ***158.09*** | | ***96.42%*** | | ***1.375*** |
| **PaSi** | | | | | |
| Rover operations[4] | 17.97 | 4.02% | 99.59% | 99.85% | 1.370 |
| Excavating | 49.78 | 11.14% | 98.86% | 99.85% | 1.542 |
| Oven robotics | 158.15 | 35.39% | 96.42% | 99.85% | 1.614 |
| Baking | 3.69 | 0.83% | 99.92% | 99.92% | 1.000 |
| Laying pavers | 46.88 | 10.49% | 98.93% | 99.85% | 1.535 |
| Grouting | 119.04 | 26.64% | 97.29% | 99.85% | 1.603 |
| Sintering | 51.32 | 11.49% | 98.82% | 99.85% | 1.545 |
| ***OVERALL SYSTEM*** | ***446.84*** | | ***90.21%*** | | ***1.481*** |
| **PaPa** | | | | | |
| Rover operations[5] | 23.74 | 1.46% | 99.45% | 99.83% | 1.414 |
| Excavating feedstock | 140.89 | 8.67% | 96.80% | 99.83% | 1.606 |
| Oven robotics | 1014.90 | 62.43% | 79.14% | 99.83% | 1.642 |
| Baking | 23.69 | 1.46% | 99.46% | 99.83% | 1.414 |
| Laying pavers | 303.33 | 18.66% | 93.25% | 99.83% | 1.628 |
| Grouting (Inner zone only) | 119.04 | 7.32% | 97.29% | 99.83% | 1.599 |
| ***OVERALL SYSTEM*** | ***1625.59*** | | ***68.75%*** | | ***1.510*** |
| **PaPo** | | | | | |
| Rover operations[4] | 17.97 | 4.53% | 99.59% | 99.82% | 1.326 |
| Excavating feedstock | 49.78 | 12.56% | 98.86% | 99.82% | 1.523 |
| Oven robotics | 158.15 | 39.91% | 96.42% | 99.82% | 1.608 |
| Baking | 3.69 | 0.93% | 99.92% | 99.92% | 1.000 |
| Laying pavers | 46.88 | 11.83% | 98.93% | 99.82% | 1.516 |
| Grouting | 119.04 | 30.04% | 97.29% | 99.82% | 1.595 |
| Polymer application | 0.77 | 0.19% | 99.98% | 99.98% | 1.000 |
| ***OVERALL SYSTEM*** | ***396.28*** | | ***91.27%*** | | ***1.354*** |
| **PaGr** | | | | | |
| Rover operations[6] | 22.76 | 4.46% | 99.48% | 99.87% | 1.454 |

| Excavating feedstock | 49.78 | 9.75% | 98.86% | 99.87% | 1.557 |
| Oven robotics | 158.15 | 30.96% | 96.42% | 99.87% | 1.619 |
| Baking | 3.69 | 0.72% | 99.92% | 99.92% | 1.000 |
| Laying pavers | 46.88 | 9.18% | 98.93% | 99.87% | 1.551 |
| Grouting | 119.04 | 23.31% | 97.29% | 99.87% | 1.609 |
| Raking and sorting rocks | 44.54 | 8.72% | 98.98% | 99.87% | 1.546 |
| Laying gravel/rock | 65.94 | 12.91% | 98.49% | 99.87% | 1.578 |
| ***OVERALL SYSTEM*** | ***510.78*** | | ***88.89%*** | | ***1.476*** |

Notes: 1. Grading and compacting both inner and outer zones. 2. Grading and compacting both inner and outer zones, hauling feedstock to make outer pavers, and hauling outer pavers for installation. 3. Grading and compacting both inner and outer zones, and hauling gravel/rocks to outer zone. 4. Grading and compacting both inner and outer, hauling feedstock to make inner pavers, and hauling inner pavers for installation. 5. Grading and compacting both inner and outer, hauling feedstock to make inner and outer pavers, and hauling inner and outer pavers for installation. 6. Grading and compacting both inner and outer, hauling feedstock to make inner pavers, hauling inner pavers for installation, and hauling gravel/rocks to outer zone.